\documentclass[12pt]{article}
\usepackage{putex}
\usepackage{graphicx}
\usepackage{latexsym,amsmath,amsfonts,amssymb,cite}
\usepackage{bbm}

\def\btab{\begin{table}[h] \begin{center} \begin{tabular}{l lp{3in}}}
      \def\etab{\end{tabular} \end{center} \end{table}}
\def\btabm{\begin{center} \begin{tabular}}
    \def\etabm{\end{tabular} \end{center}}

\def\ie{{\it i.e.}}

\def\CA{{\cal A}}

\def\CL{{\cal L}}

\def\CO{{\cal O}}

\def\BZ{\mathbb{Z}}

\begin{document}

\thispagestyle{empty}
\begin{flushright}
{\tt PUPT-2438 \\ YITP-12-47}
\end{flushright}

\begin{center}
\vspace{3cm} { \LARGE {Entanglement Entropy of a Massive Fermion on a Torus}}

\vspace{1.1cm}

Christopher P.~Herzog$^\dagger$ and Tatsuma Nishioka$^\ddagger$

\vspace{0.8cm}

{\it $\dagger$ C.~N.~Yang Institute for Theoretical Physics \\ Stony Brook University, Stony Brook, NY  11794, USA}

\vspace{0.2cm}
{\it $\ddagger$ Department of Physics, Princeton University, Princeton, NJ 08544, USA}

\vspace{2cm}

\end{center}

\begin{abstract}
\noindent
The R\'enyi entropies of a massless Dirac fermion on a circle with chemical potential 
are calculated analytically at nonzero temperature by using the bosonization method.
The bosonization of a massive Dirac fermion to the sine-Gordon model lets us obtain the small mass corrections to the entropies.
We numerically compute the R\'enyi entropies by putting a massive fermion on the lattice and find agreement between the analytic and numerical results.
In the presence of a mass gap, 
we show that corrections to R\'enyi and entanglement entropies in the limit $m_{\rm gap} \gg T$ scale as $e^{-m_{\rm gap}/T}$.  We
also show that when there is ground state degeneracy in the gapless case, the limits $m_{\rm gap} \to 0$ and $T \to 0$ do not commute. \end{abstract}

\vspace{4cm}

\noindent
January 2013

\pagebreak
\setcounter{page}{1}

\tableofcontents

\section{Introduction}

Entanglement entropy is a unifying theme in many different areas of theoretical physics today.
In relativistic field theories, certain special kinds of entanglement entropy show monotonicity properties under renormalization group flow \cite{Casini:2006es,Casini:2012ei}.  For conformal field theories in $(1+1)$-dimensions, numerical computation of the entanglement entropy provides a rapid way to calculate the central charge $c$.  In the context of condensed matter physics, entanglement entropy can detect exotic phase transitions for systems lacking a local order parameter.
The Ryu-Takayanagi proposal \cite{Ryu:2006bv,Ryu:2006ef}  for computing the entanglement entropy holographically connects this circle of ideas to general relativity and string theory via the AdS/CFT correspondence \cite{Maldacena:1997re,Gubser:1998bc,Witten:1998qj}.
See refs.\ \cite{Calabrese:2009qy,EislerPeschel,Casini:2009sr,Nishioka:2009un,Takayanagi:2012kg} for reviews.

Recall the entanglement entropy is defined from a reduced density matrix $\rho_A$.  We start by partitioning the Hilbert space into pieces $A$ and complement $\bar A = B$.  Typically $A$ corresponds to a spatial region.  We form the reduced density matrix $\rho_A = \tr_B \rho$ by tracing over the degrees of freedom in $B$.  Finally, the entanglement entropy is defined to be
\begin{equation}
\label{SEdef}
S \equiv - \tr \rho_A \log \rho_A \ .
\end{equation}
Related quantities are the R\'enyi entropies
\begin{equation}
\label{Renyidef}
S_n = \frac{1}{1-n} \log \tr (\rho_A)^n \ .
\end{equation}
Note that the entanglement entropy can be determined from the limit $S = \lim_{n \to 1} S_n$. 

Given entanglement entropy's prominent role, it is surprisingly difficult to compute, even for free field theories in $(1+1)$-dimensions.  In this paper, we present some new results for the free, massive, Dirac fermion in $(1+1)$-dimensions.
We are particularly interested in thermal and finite size corrections to the entanglement entropy, and so we place our massive fermion on a torus.\footnote{%
A prequel \cite{Herzog:2012bw} to this paper considered the free massive scalar on a torus; the two papers can be read independently.
}
  We allow for a nonzero chemical potential as well.

Before proceeding further, let us briefly review the known results for the free fermion.
Consider a massless Dirac fermion on the real line where the region $A$ consists of $p$ intervals whose endpoints are described by the pairs of numbers $(x_a, y_a)$, $a = 1, \ldots, p$.  In this case, the entanglement entropy takes the form \cite{Casini:2005rm}\footnote{%
See also ref.\ \cite{Calabrese:2004eu}.
} 
\begin{equation}
S = \frac{1}{3} \log \left| \frac{\prod_{a,b} (x_a-y_b)}{ \epsilon^p \prod_{a<b}(x_a-x_b)(y_a-y_b)}  \right| + c_0 \ ,
\label{Smulti}
\end{equation}
where $\epsilon$ is a UV cutoff and $c_0$ is a cutoff dependent constant.
Given that the fermion is massless, we can use conformal symmetry to map the plane to a cylinder with either time or space 
compactified  \cite{Calabrese:2004eu}.  
In the first case, we make the replacement  $(x-y) \to \sinh \pi T(x-y)$ where $T$ is the temperature.
In the second case, we instead send  $(x - y) \to \sin \pi (x-y)/L$ where $L$ is the circumference of the spatial circle.

If we turn on a mass $m>0$, we lose conformal symmetry, and the computations get correspondingly more difficult.  
For the fermion on the real line at zero temperature and a single interval, the entanglement entropy can be expressed in terms of a solution to the Painlev\'e V equation \cite{Casini:2005rm}.  While in general only a numerical solution to the differential equation is available, one can find small and large mass expansions.  For small mass, the leading log result is
\begin{equation}
\label{Ssmallmsingle}
S = \frac{1}{3} \log \frac{\ell}{\epsilon} - \frac{1}{6} (m \ell \log m \ell)^2 + O((m \ell)^2 \log m \ell) \ .
\end{equation}
At large mass, there is instead exponential suppression:\footnote{%
See also refs.\ \cite{Cardy:2007mb,Doyon:2008vu}.
}
\begin{equation}
S \sim \frac{1}{8} \sqrt{\frac{\pi}{m \ell}} e^{-2 m \ell} \ .
\end{equation}
For multiple intervals, ref.\ \cite{Casini:2009vk} provides a small mass expansion.  
The leading log correction is instead $-\frac{1}{6} (m \ell_t \log m \epsilon)^2$ where 
$\ell_t$ is the total length of all of the intervals in $A$. 

This paper contains two principal results.  The first is a computation of the R\'enyi entropies for a region $A$ consisting of multiple intervals  where both $1/T$ and $L$ are kept finite, the chemical potential $\mu$ can be different from zero, but $m=0$.  
Previously, only the single interval R\'enyi entropy was available \cite{Azeyanagi:2007bj,Ogawa:2011bz}.  
We compute the entanglement entropy from the R\'enyi entropies by analytic continuation.  While the R\'enyi entropies are expressed compactly in terms of elliptic theta functions, 
our entanglement entropy is given as an infinite sum.  
In the limit where $L \to \infty$ or 
$T\to 0$, only the first few terms in the sum contribute, and we recover the cylinder version of 
eq.\ (\ref{Smulti}). However, there is a subtlety in the 
$T\to 0$ limit that we will return to shortly.

The second principal result in this paper is a computation of the leading small mass correction to the entanglement entropy 
when both $1/T$ and $L$ are finite. The correction is obtained using the equivalence of the massive Dirac fermion to the sine-Gordon model \cite{Coleman:1974bu,Mandelstam:1975hb}.  The result is expressed as a double integral over a product of elliptic theta functions.  We are able to perform the double integral numerically and match the result to a numerical lattice computation of the entanglement entropy.
We can also perform the integral in the limit 
$1/T, L \to \infty$ where we recover the multi-interval version of the 
leading log correction (\ref{Ssmallmsingle}).
 
We have found several interesting features of the entanglement entropy in the small mass and low temperature regime.
The first is that the limits $m\to 0$ and 
$T\to 0$ of the entanglement entropy do not commute when there is ground state degeneracy.
In particular, when the massless Dirac fermions have periodic boundary conditions around the circle, the ground state of the system is four-fold degenerate.  If we first set $m=0$ and then take 
$T\to 0$, the system is not in a pure state.  However, if we instead send 
$T\to 0$ and then take $m \to 0$, the system will be pure.
For pure states, the entanglement entropy of a region and its complement must be the same, $S(A) = S(\bar A)$. 
For example, consider the single interval case, $\ell = x-y$, of eq.\ (\ref{Smulti}) where we have conformally mapped to the cylinder $(x-y) \to \sin \pi (x-y)/L$.  Then the entanglement entropy takes the form
\begin{equation}
S = \frac{1}{3} \log \left( \frac{L}{\pi \epsilon} \sin \frac{\pi \ell}{L} \right) + c_0 \ ,
\label{SL}
\end{equation}
which clearly satisfies $S(\ell) = S(L-\ell)$.
If instead we take $m \to 0$ first, we find a correction to eq.\ (\ref{SL}) that reflects the ground state degeneracy.  
There is a similar correction in the multi-interval case.

For Dirac fermions with antiperiodic boundary conditions, there is a unique ground state.  The mass gap to the first excited state is $m_{\rm gap} = \pi / L$.  In this case, we can examine the corrections to the entanglement entropy in the low temperature 
limit $T \ll \pi/L$.  
The prequel \cite{Herzog:2012bw} to this paper, based on an investigation of the massive (1+1)-dimensional scalar, 
conjectured that such corrections should be exponentially suppressed.
Indeed, we are able to confirm this conjecture for the antiperiodic fermions:
\begin{equation}
S(T) - S(0) \sim e^{- m_{\rm gap}/T}.
\end{equation}
Moreover, our result for the R\'enyi entropies allows us to determine the coefficient in front of the exponential factor.
For the periodic fermions with $m>0$, we find similar behavior, but in this case, our results come from a numerical lattice computation.

The Lagrangian density for a Dirac fermion in what would be mostly minus signature for the metric if we had more than one spatial direction is
\begin{equation}
{\mathcal L}_{DF} = \bar \Psi (i\gamma^\mu \partial_\mu -m )  \Psi  \ .
\label{LDF}
\end{equation}
Our conventions for the gamma matrices are that $\{\gamma^\mu, \gamma^\nu \} = 2\eta^{\mu\nu} = 2(+-)$.  We define $\bar \Psi \equiv \Psi^\dagger \gamma^0$.
We choose gamma matrices $\gamma^0 = \sigma^1$ and $\gamma^1 = -i \sigma^2$.

\section{Bosonization and Conformal Field Theory}\label{sec:Massless}
Following \cite{Casini:2005rm}, we consider the entanglement entropy of a free Dirac fermion on a torus with multiple intervals $(u_a, v_a)~(a=1,\dots,p)$.
Instead of having a single field on the $n$-covering space, we introduce $n$ decoupled fields $\tilde\Psi^k ~ (k=-\frac{n-1}{2}, \cdots, \frac{n-1}{2})$ living on a single torus. They are multivalued around the branch points $u_a, v_a$ around which they get phases $e^{i\frac{ 2\pi k}{n}}$ and $e^{-i\frac{ 2\pi k}{n}}$, respectively.
Here we define the single-valued field $\Psi^k$ by introducing an external gauge field $\tilde \Psi^k(x)= e^{i \int_{x_0}^x dx'^\mu \CA_\mu^k(x')} \Psi^k (x)$. It follows that the Lagrangian is given by
\begin{align}
	\CL_k = i \bar \Psi^k \gamma^\mu (\partial_\mu + i \CA_\mu^k) \Psi^k - m \bar \Psi^k \Psi^k \ .
\end{align}
The gauge field is almost pure gauge except at the branch points where delta function singularities are necessary
to recover the correct phases of the multivalued fields
\begin{align}
	\epsilon^{\mu\nu}\partial_\nu \CA_\mu^k(x) = \frac{2\pi k}{n} \sum_{a=1}^p \left[ \delta^{(2)}(x - u_a) - \delta^{(2)}(x - v_a) \right] \ .
\end{align}
The partition function is then obtained in a factorized form
\begin{align}\label{FermPart}
	Z[n] = \prod_{k=-\frac{n-1}{2}}^{\frac{n-1}{2}}Z_k \ ,
\end{align}
where $Z_k$ is the partition function of the $k$-th fermion coupled to the external gauge field $\CA^k_\mu$
\begin{align}\label{Zk}
	Z_k = \langle e^{i\int \CA^k_\mu j^\mu_k d^2 x} \rangle \ ,
\end{align}
with the current $j^\mu_k = \bar\Psi^k \gamma^\mu \Psi^k$.

In $(1+1)$-dimensions, one can describe fermions in terms of non-local operators of scalars. 
The current is mapped to the derivative of a scalar field
\begin{align}\label{bosonize}
	j_k^\mu \to \frac{1}{2\pi} \epsilon^{\mu\nu}\partial_\nu \phi_k \ ,
\end{align}
and the Lagrangian of the $k$-th fermion becomes that of $k$-th real free massless scalar field $\phi$: $\CL_k = \frac{1}{8\pi} \partial_\mu\phi_k \partial^\mu \phi_k$.
It follows from \eqref{Zk} and \eqref{bosonize} that $Z_k$ can be written as the correlation function of the vertex operators
\begin{align}
	Z_k = \langle \prod_{a=1}^p V_{k}(u_a) V_{-k}(v_a) \rangle\ ,
\end{align}
where the vertex operator $V_k$ is defined as $V_k(x) = e^{-i \frac{k}{n} \phi_k(x)}$.

The scalar field is a compactified boson with radius $R = 2$ so as to reproduce
the partition function of a Dirac fermion on a torus.\footnote{%
In our conventions, $R=\sqrt{2}$ is the self-dual radius.
} 
We have used the bosonization technique without specifying the spin structure of the fermion on a torus.
We shall be more careful to distinguish the spin structures in the following.
\begin{table}
\[
\begin{array}{|c|c|c|}
\hline
\nu & \mbox{sector} & (\nu_1, \nu_2) \\
\hline
1 & \mbox{(R,R)} & (0,0) \\
2 & \mbox{(R,NS)} & (0,\frac{1}{2}) \\
3 & \mbox{(NS,NS)} & (\frac{1}{2}, \frac{1}{2}) \\
4 & \mbox{(NS,R)} & (\frac{1}{2}, 0) \\
\hline
\end{array}
\]
\caption{
Conventions for fermion boundary conditions.
\label{fig:spinsector}
}
\end{table}
The torus is specified by two periods which we take as $1$ and $\tau = i\beta$ where $\beta = 1/(TL)$ is the dimensionless inverse temperature.\footnote{We rescale the spacetime coordinates by $L$. }
Let $z$ be a holomorphic coordinate on the torus; then it has the periodicity $z\sim z + 1$ and $z \sim z + \tau$.
The holomorphic part of the fermion on the torus satisfies four possible boundary conditions
\begin{align}\label{FermBC}
\psi(z+1) = e^{2i\pi \nu_1}\psi(z) \ ,\qquad \psi (z+\tau) = e^{2i\pi \nu_2}\psi(z) \ ,
\end{align}
where $\nu_1$ and $\nu_2$ take $0$ or $\frac{1}{2}$. The anti-holomorphic part satisfies the same boundary conditions as the holomorphic part.
We denote the $\nu=(\nu_1,\nu_2)$ sector where $\nu = 1,2,3,4$ correspond to $(0,0),(0,1/2), (1/2,1/2), (1/2,0)$, respectively
(see Table~\ref{fig:spinsector}).
The corresponding partition function $Z_\nu$ is given by
\begin{align}
	Z_\nu & = \frac{1}{2} \bigg| \frac{\vartheta_\nu(0|\tau)}{\eta(\tau)}\bigg|^2 \ .
\end{align}
Corresponding to the sector $\nu$, we can find a boson whose partition function 
agrees with $Z_\nu$ in the fermionic theory.
The correlation function of the vertex operators on the torus in the $\nu$ sector is given in 
ref.\ \cite{DiFrancesco:1997nk}
\begin{align}\label{VertexCorr}
	\langle \CO_{e_1}(z_1,\bar z_1) \cdots \CO_{e_N}(z_N,\bar z_N) \rangle_\nu = \prod_{i<j}\left| \frac{\partial_z \vartheta_1(0|\tau)}{\vartheta(z_j - z_i|\tau)} \right|^{-\frac{e_i e_j}{2}} \left| \frac{\vartheta_\nu (\frac{\sum_{i} e_i z_i}{2} | \tau)}{\vartheta_\nu (0|\tau)} \right|^2 \ ,
\end{align}
where $R=2$ and $\CO_e$ is a vertex operator defined by $\CO_e (z,\bar z) = e^{i \frac{e}{2}\phi(z,\bar z)}$.
It follows from this formula that 
\begin{align}\label{PFknu}
	  Z_{k,\nu} &= \Bigg| \frac{\prod_{a<b}  \vartheta_1(u_a-u_b|\tau )\, \vartheta_1(v_a-v_b|\tau ) }{ \prod_{a,b}  \vartheta_1(u_a-v_b|\tau )} \cdot (\epsilon\,\partial_z \vartheta_1(0|\tau))^p \Bigg|^\frac{2k^2}{n^2} \cdot  \Bigg| \frac{\vartheta_{\nu}(\frac{k}{n}\sum_a (u_a - v_a)|\tau)}{\vartheta_{\nu}(0|\tau)}\Bigg|^2 \ ,
\end{align}
where we denote the partition function of the $k$-th fermion in the $\nu$ sector by $Z_{k,\nu}$.
We normalize the partition function such that $Z_{k,\nu} = 1$ when there are no branch points.
Since the theta function behaves as $\vartheta_1(z|\tau) \sim z$ in the small $z$ limit, we put the UV cutoff $\epsilon$ to split the coincident points.\footnote{Here $\epsilon$ is dimensionless. The dimensionful UV cutoff is $\epsilon L$.}
Finally the total partition function \eqref{FermPart} in the $\nu$ sector is obtained as
\begin{align}
	\log Z_\nu [n] &= \frac{n^2 - 1}{6n} \log \Bigg| \frac{\prod_{a<b}  \vartheta_1(u_a-u_b|\tau )\, \vartheta_1(v_a-v_b|\tau ) }{ \prod_{a,b}  \vartheta_1(u_a-v_b|\tau )} \cdot (\epsilon\,\partial_z \vartheta_1(0|\tau))^p \Bigg| \nonumber \\
	& \qquad + \sum_{k=-\frac{n-1}{2}}^{\frac{n-1}{2}} 2\log\Bigg| \frac{\vartheta_{\nu}(\frac{k}{n}\sum_a (u_a - v_a)|\tau)}{\vartheta_{\nu}(0|\tau)}\Bigg| \ .
\end{align}
The R\'enyi entropy has the following form
\begin{align}
\label{Renyi}
	S_{n}^{(\nu)} &= \frac{1}{1-n}\left( \log Z_\nu[n] - n \log Z_\nu [1] \right) \nonumber\\
	&= S_{n,0} + S_{n,1}^{(\nu)} \ .
\end{align}
Here the first term is universal,
\begin{equation}\label{REuniv}
S_{n,0} = - \frac{n+1}{6n}\log \Bigg| \frac{\prod_{a<b}  \vartheta_1(u_a-u_b|\tau )\, \vartheta_1(v_a-v_b|\tau ) }{ \prod_{a,b}  \vartheta_1(u_a-v_b|\tau )} \cdot (\epsilon\,\partial_z \vartheta_1(0|\tau))^p \Bigg| \ ,
\end{equation}
and the second depends on the spin structure,
\begin{equation}\label{REspin}
S_{n,1}^{(\nu)} =\frac{2}{1-n}\sum_{k=-\frac{n-1}{2}}^{\frac{n-1}{2}} \log\Bigg| \frac{\vartheta_{\nu}(\frac{k}{n}\sum_a (u_a - v_a)|\tau)}{\vartheta_{\nu}(0|\tau)}\Bigg| \ .
\end{equation}
Note that the R\'enyi entropies in the $\nu=1$ sector are divergent because of the $\theta_1(0|\tau)$ inside the logarithm in $S_{n,1}^{(1)}$.  We will have little to say about the massless $\nu=1$ sector in what follows.

\subsection{Adding chemical potential} \label{sec:ACP}

In Lorentzian signature, a chemical potential is equivalent to introducing a constant time-like component of the vector potential $A_t = \mu$.  In Euclidean signature, the chemical potential becomes pure imaginary, $A_{t_E} = i \mu$.  These considerations suggest that we can understand the dependence of entanglement entropy on chemical potential by thinking about flat gauge connections on the torus.

Let the vector potential be ${\mathcal A}^k = a_t \, dt_E +a_x \, dx +\ldots$ where $a_t$ and $a_x$ are constant and the ellipsis denotes terms responsible for the twisted boundary conditions around $u_a$ and $v_b$.  Note that $a_t$ and $a_x$ are defined only up to gauge transformations which shift ${\mathcal A}^k  \to {\mathcal A}^k  + 2 \pi n ( \frac{1}{\beta} dt_E + dx)$ where $n$ is an integer.
Such a flat connection contributes to the partition function through eq.\ (\ref{Zk}).

In the bosonized picture, 
the periodic scalar has boundary conditions along the thermal and spatial circles that are characterized by two winding numbers $(w, w')$:
\begin{align}
	\phi(z+1) = \phi(z) + 2\pi R \,w \ , \qquad \phi(z+\tau) = \phi(z) + 2\pi R \, w' \ .
\end{align}
The expectation value of the vertex operators is then computed by summing those over the topological sectors:
\begin{align}
	\left \langle \prod_j {\mathcal O}_{e_j} (z_j , \bar z_j) \right \rangle = \left. \sum_{w, w' \in \BZ} \left \langle \prod_j {\mathcal O}_{e_j} (z_j , \bar z_j) \right \rangle \right|_{(w, w')} \,
	e^{2 i(\beta a_t w - a_x w')} \ .
\end{align}
When $\tau = i\beta$, the $(w, w')$ sector is related to the $(0,0)$ sector (see \cite{DiFrancesco:1997nk}):
\begin{align}
	\left. \left \langle \prod_j {\mathcal O}_{e_j} (z_j , \bar z_j)  \right \rangle \right|_{(w, w')}  = 
	\left. \left \langle \prod_j {\mathcal O}_{e_j} (z_j , \bar z_j)  \right \rangle \right|_{(0,0)}
	 \exp \left[ 2 \pi i  \sum_j e_j\left( \frac{ \mbox{Im}(z_j)}{\beta} w' + \mbox{Re}(z_j) w \right) \right] 
	 \ .
\end{align}
From this result, we see that the effect of the flat gauge connection can be incorporated in the correlation function by making the shift
\begin{equation}
\sum_j \frac{e_j}{2} z_j \to \sum_j \frac{e_j}{2} z_j + \frac{\beta}{2 \pi} ( a_t - i a_x) \ 
\label{zshift}
\end{equation}
in the $\nu$ dependent portion of the correlation function.

Alternately, through the relation $\tilde \Psi^k(x)= e^{i \int_{x_0}^x dx'^\mu \CA_\mu^k(x')} \Psi^k (x)$, we can trade the flat gauge connection for a shift in boundary conditions.  From this expression, one may make the identifications $a_x = \pm 2 \pi \nu_1$ and $a_t = \pm 2 \pi \nu_2 / \beta$.   
Indeed, in eq.\ (\ref{VertexCorr}), we can rewrite the $\nu$ dependent term with the use of the formulae \eqref{ShiftTheta} in appendix \ref{app:theta} as
\begin{align}
\left| \frac{ \vartheta_\nu ( \frac{\sum_i e_i z_i}{2} | \tau ) }{\vartheta_\nu(0|\tau)} \right|^2
&=
\left| \frac{ \vartheta_1 ( \frac{\sum_i e_i z_i}{2} - \nu_1 \tau - \nu_2 | \tau )}{\vartheta_1(-\nu_1 \tau - \nu_2| \tau)} \right|^2 \ , \nonumber\\
&=
\left| \frac{ \vartheta_1 ( \frac{\sum_i e_i z_i}{2} + \frac{\beta}{2\pi}(a_t - i a_x) | \tau )}{\vartheta_1(\frac{\beta}{2\pi}(a_t - i a_x)| \tau)} \right|^2 \ .
\label{atax}
\end{align}
To match the shift (\ref{zshift}), we made the sign choices $a_x = 2 \pi \nu_1$ and $a_t = - 2 \pi \nu_2 / \beta$.  

Introducing a chemical potential, by analytic continuation, is equivalent to introducing an imaginary $a_t = i \mu$. 
From the structure of eq.\ (\ref{zshift}), it is clear that adding a chemical potential is also equivalent to introducing a real $a_x$.
This second equivalence makes it clear that the effect of chemical potential must be periodic with period $2 \pi$.  
Restoring dimensions, we see that the periodicity $2 \pi /L$ is precisely the energy level spacing on the torus.

 From eq.\ (\ref{atax}), one then obtains the partition function of the $k$-th fermion in the $\nu$ sector with chemical potential 
\begin{align}
	  Z_{k,\nu} &= \Bigg| \frac{\prod_{a<b}  \vartheta_1(u_a-u_b|i\beta )\, \vartheta_1(v_a-v_b|i\beta ) }{ \prod_{a,b}  \vartheta_1(u_a-v_b|i\beta )} \cdot (\epsilon\,\partial_z \vartheta_1(0|i\beta))^p \Bigg|^\frac{2k^2}{n^2} \cdot  \Bigg| \frac{\vartheta_{\nu}(\frac{k}{n} \frac{\ell_t}{L} + \frac{i\beta \mu}{2\pi}|i\beta)}{\vartheta_{\nu}(\frac{i\beta \mu}{2\pi}|i\beta)}\Bigg|^2 \ ,
\end{align}
where $\ell_t = L \sum_a(v_a - u_a)$ is the total width of the intervals.
The universal part of the R\'enyi entropy remains the same \eqref{REuniv} and 
the part depending on the spin structure \eqref{REspin} is altered to
\begin{equation}
S_{n,1}^{(\nu)} =\frac{2}{1-n}\sum_{k=-\frac{n-1}{2}}^{\frac{n-1}{2}} \log\Bigg| \frac{\vartheta_{\nu}(\frac{k}{n} \frac{\ell_t}{L} + \frac{i\beta \mu}{2\pi}|i\beta)}{\vartheta_{\nu}(\frac{i\beta \mu}{2\pi}|i\beta)}\Bigg| \ .
\end{equation}
Our result reduces to that of ref.\,\cite{Ogawa:2011bz} for one interval.

\subsection{Low temperature limit}
In this section, we consider a series expansion of $S_n^{(\nu)}$ in the low temperature limit,
$
\tau = i \beta \to i \infty
$.
We take advantage of the product representation of the theta functions (see appendix \ref{app:nonthermal}).
The universal term $S_{n,0}$ on the right hand side of (\ref{Renyi}) becomes 
\begin{equation}
\label{Snzero}
S_{n,0} = - \frac{n+1}{6n} \log \left|
\frac{
\prod_{a<b} \sin \pi(u_a-u_b) \sin \pi (v_a - v_b)
}
{\prod_{a,b} \sin \pi(u_a - v_b) } (\pi \epsilon)^p
\right|
 + O(e^{-2 \pi \beta}) \ ,
\end{equation}
in this limit.  Note the entanglement entropy contribution can be straightforwardly recovered by setting $n=1$, which in turn agrees with the spatial cylinder version of eq.\ (\ref{Smulti}) reviewed in the introduction.  

However, to claim complete agreement with eq.\ (\ref{Smulti}), we need to check that $S_{n,1}^{(\nu)}$ does not contribute at zero temperature.
Consider low temperature expansions of $S_{n,1}^{(\nu)}$ for the spin structures $\nu = 2$ and 3 corresponding to thermal boundary conditions.
(The non-thermal spin structures $\nu=1$ and 4 are given in appendix \ref{app:nonthermal}.)  
For $\nu=2$, defining $r \equiv \sum_a (v_a - u_a)$, we find that
\begin{equation}\label{S2delta}
S_{n,1}^{(2)} = \delta s(n,r) + s_2(n,r) \ ,
\end{equation}
where  
\begin{equation}
\delta s(n,r) = \frac{2}{1-n} \sum_{k = -\frac{n-1}{2}}^{\frac{n-1}{2}} \log \left| \cos \frac{ \pi k r}{n} \right| \ ,
\end{equation}
and
\begin{equation}
s_2(n,r) = \frac{4}{1-n} \sum_{j=antiperiodic1}^\infty \frac{(-1)^{j+1}}{j} \frac{1}{e^{2 \pi  \beta j} - 1} \left( \frac{ \sin ( \pi j r)}{\sin \left( \frac{\pi j r}{n} \right) } - n \right) \ .
\end{equation}
For $\nu=3$, we find instead that 
\begin{align}
\label{Sn13}
S_{n,1}^{(3)} &= \frac{2}{1-n} \sum_{j=1}^\infty \frac{(-1)^{j+1}}{j\sinh \pi \beta j} \left( \frac{ \sin(\pi j r)}{\sin \left( \frac{\pi j r}{n} \right)} - n \right)  \ .
\end{align}
Thus for spatially antiperiodic fermions, eq.\ (\ref{Snzero}) is the whole story at zero temperature, while spatially periodic fermions get an extra correction $\delta s(n,r)$.

To investigate the entanglement entropy, we take the $n\to 1$ limit.
Much of this limit is straightforward:
\begin{align}
\lim_{n\to 1} s_2(n,r) &= 2 \sum_{j=1}^\infty \frac{(-1)^{j+1}}{j} \frac{ 1 - \pi j r \cot (\pi j r)}{\sinh \pi \beta j}e^{-\pi \beta j}  \ , \\
\lim_{n\to 1} S_{n,1}^{(3)} &= 2 \sum_{j=1}^\infty \frac{(-1)^{j+1}}{j} \frac{1 - \pi j r \cot (\pi j r)}{\sinh \pi \beta j}   \ .
\label{S113}
\end{align}
These contributions vanish exponentially in the $\beta \to \infty$ limit.
Our analysis of $\delta s(n,r)$ is incomplete. 
 We find that 
\begin{equation}
\delta s(n,1) = 2 \ln 2 \ ,
\end{equation}
 for all $n$, consistent with the fact that spatially periodic Dirac fermions have a ground state degeneracy equal to four.  For small $r$, we were able to obtain an asymptotic Euler-Maclaurin type expansion:
\begin{equation}
\lim_{n \to 1} \delta s(n,r) = 2 \sum_{j=1}^\infty \frac{(2^{2j}-1) }{j} B_{2j} \, \zeta(2j) \, r^{2j} \ ,
\end{equation}
where $B_{j}$ is a Bernoulli number and $\zeta(x)$ is the Riemann zeta function.
Unfortunately, this expression is not Borel summable.

\subsection{High temperature expansion}
To investigate high temperature behavior, 
we use the modular transformation rules for the theta functions: 
\begin{align}
	\vartheta_1(z|\tau) = -(-i \tau)^{-1/2} e^{-\pi iz^2/\tau} \vartheta_1 (z/\tau | {-}1/\tau) \ .
\end{align}
The modular transformations of the other theta functions are given in appendix \ref{app:theta}.
The asymptotic form of the theta function depends on the value of $z$ in the small $\beta$ limit:
\begin{align}
	\vartheta_1 (z/\tau | {-}1/\tau) = -2i \, e^{-\frac{\pi}{4\beta}} \sinh \frac{\pi z}{\beta} + O( e^{\frac{3\pi}{\beta} (z-3/4)}) \ , \qquad \left(0\le z \le \frac{1}{2}\right)\ ,
\end{align}
where $\tau = i \beta$ was used. For $1/2 \le z \le 1$, one may use the periodicity of the theta function:
\begin{align}\label{ShiftTheta1}
	\vartheta_1 (z/\tau | {-}1/\tau) =  e^{\frac{\pi i}{\tau}(2z-1)} \vartheta_1 ((1-z)/\tau | {-}1/\tau) \ .
\end{align}
When $v_p - u_1\le 1/2$, the leading term of the universal part $S_{n,0}$ of the R\'enyi entropy can be written
\begin{align}
\label{Sn0highT}
S_{n,0} &=
-\frac{(1+n)}{6n} \Biggl( \frac{\pi r^2}{\beta} + 
 \ln\left| \frac{ \prod_{a<b} \sinh \frac{\pi(u_a - u_b)}{\beta} \sinh \frac{\pi (v_a - v_b)}{\beta}}{\prod_{a,b} \sinh \frac{ \pi (u_a-v_b)}{\beta}}\left(\frac{\pi \epsilon}{\beta} \right)^p \right|
 \Biggr) \\
 &  \qquad + O\left( e^{\frac{2\pi }{\beta} (v_p-u_1 - 1)} \right) \ . \nonumber
 \end{align}
For $\nu=2$ and 3, we find that
\begin{align}
S_{n,1}^{(\nu)} &= \frac{(1+n)}{6n} \frac{\pi r^2}{\beta} - \frac{2}{1-n} \sum_{j=1}^\infty \frac{(-1)^{\nu j}}{j}\frac{1}{\sinh \frac{\pi j}{\beta}} \left( \frac{\sinh \frac{\pi j r}{\beta}}{\sinh\frac{\pi j r}{n\beta}} -n\right) \ . 
\end{align}
The entanglement entropy limit is given by
\begin{align}
\lim_{n\to 1} S_{n,1}^{(\nu)}  &=  \frac{\pi r^2}{3\beta} - 2 \sum_{j=1}^\infty \frac{(-1)^{\nu j}}{j}\frac{1 - \frac{\pi j r}{\beta} \coth\left( \frac{\pi j r}{\beta}\right) }{\sinh \frac{\pi j}{\beta}} \ .
\end{align}
Similar results for $\nu=1$ and 4 are given in appendix \ref{app:nonthermal}.

Note that the leading $\pi r^2 / \beta$ dependence cancels between $S_{n,0}$ and $S_{n,1}^{(\nu)}$.  
To recover the temporal cylinder version of eq.\ (\ref{Smulti}), we need to take $\beta \to 0$ while keeping $u_a/\beta$ and $v_b/\beta$ fixed.  

\subsection{Mutual information}\label{sec:MI}
The mutual R\'enyi information is an important measure of the entanglement between two intervals.
Given two intervals $A$ and $B$ of length $\ell_1$ and $\ell_2$ separated by $\ell_3$ on a circle of circumference $L$, the mutual R\'enyi information is
\begin{align}
	I_n(A,B) = S_n(A) + S_n(B) - S_n(A\cup B) \ .
\end{align}
The definition makes clear that the mutual information is free of UV divergences, unlike the entanglement entropy.
Using eq.\ (\ref{Renyi}), the mutual R\'enyi information of two intervals for a massless Dirac fermion on a circle at finite temperature becomes
\begin{align}\label{RMI}
	I_n(A,B) =& \,\frac{n+1}{6n} \log\left| \frac{\vartheta_1(\frac{\ell_1 + \ell_3}{L}|\tau) \vartheta_1(\frac{\ell_2 + \ell_3}{L}|\tau)}{\vartheta_1(\frac{\ell_1 + \ell_2 + \ell_3}{L}|\tau)\vartheta_1(\frac{\ell_3}{L}|\tau)} \right| \nonumber\\
	&\qquad - \frac{2}{1-n} \sum_{k=- \frac{n-1}{2}}^{\frac{n-1}{2}} \log \left|
	\frac{\vartheta_\nu( \frac{k}{n}\frac{\ell_1 + \ell_2}{L}|\tau) \vartheta_\nu(0|\tau)}{\vartheta_\nu (\frac{k}{n}\frac{\ell_1}{L} |\tau) \vartheta_\nu (\frac{k}{n}\frac{\ell_2}{L} |\tau)} \right| \ .
\end{align}

\begin{figure}[htbp]
	\centering
	\includegraphics[width=5.5cm]{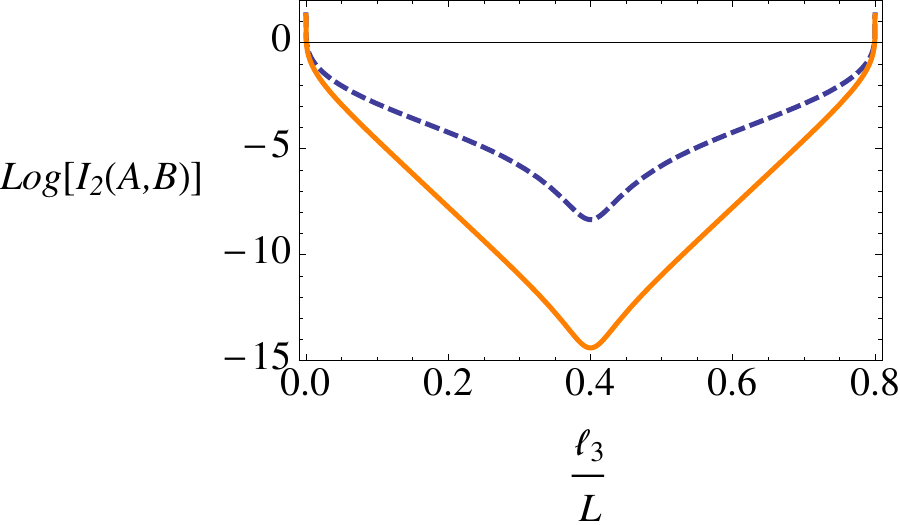}~ 
	\includegraphics[width=5.5cm]{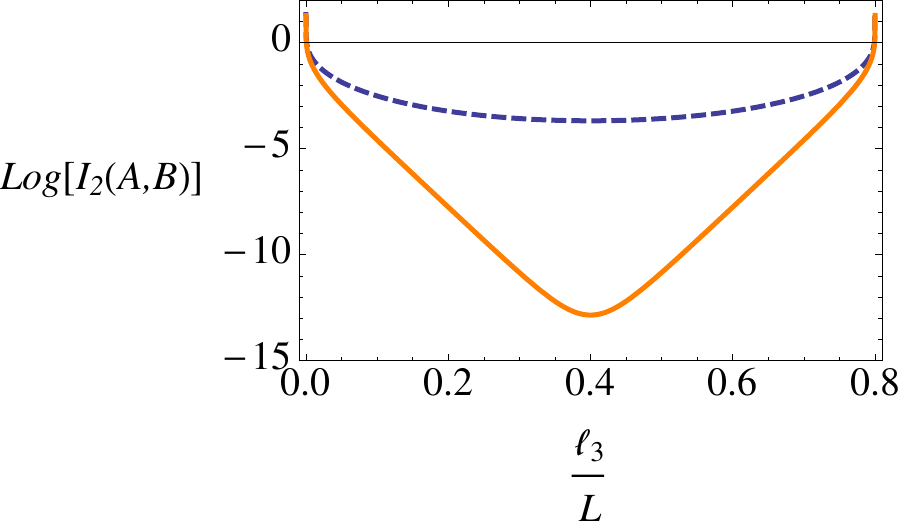}~
	\includegraphics[width=5.5cm]{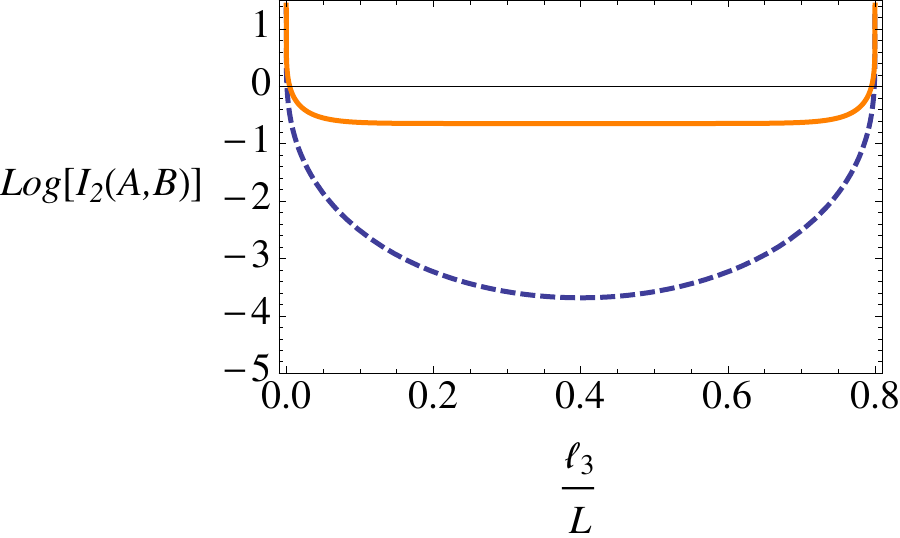}
	\caption{The mutual R\'enyi informations of two intervals $A$ and $B$ of width $\ell_1 = \ell_2 = L/10$ with $n=2$ in the $\nu=2$ [Left], $\nu=3$ [Middle] and $\nu=4$ [Right] sectors. $\ell_3$ is the distance between the two intervals. The blue dashed and orange solid curves are for $\beta = 10,1/5$, respectively.}
	\label{LogMI}
\end{figure}

The logarithmic plots of the mutual R\'enyi informations for $n=2$  in the $\nu=2,3,4$ sectors are shown in Fig.~\ref{LogMI}. 
The mutual information is completely finite and positive. 
We let the width of two intervals $A$ and $B$ be $\ell_1 = \ell_2 = L/10$ and plot the mutual information with respect to the distance $\ell_3$ between them. 
Since the two intervals are on a circle of radius $L$, $I_2$ is symmetric under $\ell_3 \to L-\ell_1 - \ell_2 - \ell_3$ as is clear from the expression \eqref{RMI}.
The plots for $n \geq 3$ are qualitatively similar.

We can use the high and low temperature expansions of $S_n^{(\nu)}$ described above to get a better understanding of the behavior of $I_n$.  At large $T$, the theta functions can be replaced by hyperbolic sine functions, as in the expansion (\ref{Sn0highT}).  
In the $\nu=2$ and 3 cases, 
expanding the hyperbolic sines, for $\ell_3 < L/2$ we find $I_2 \sim e^{-2 \pi \ell_3 T}$, while for $\ell_3 > L/2$, by symmetry, $I_2 \sim e^{-2 \pi (L - \ell_1 - \ell_2 - \ell_3)T}$.  
The $\nu=4$ spin structure, however, develops an order one contribution to the entanglement entropy at high $T$, as can be seen from the expansion (\ref{Sn14S}).
At low $T$, the theta functions are replaced by sine functions, as in the expansion (\ref{Snzero}).  However, for $\nu=2$, there is an extra contribution from $\delta s(n,r)$ because of the ground state degeneracy. 

Before closing this section, we compare our findings to the holographic computation \cite{Headrick:2010zt,Fischler:2012uv} where the mutual information undergoes a phase transition as the distance between the two intervals increases, \ie, $I(A,B)\neq 0$ for small $\ell_3$ while $I(A,B)=0$ for large $\ell_3$.
In our case, the finite volume and finite number of degrees of freedom prevent a phase transition from happening.  However, for large temperatures the mutual information exponentially falls off as $\ell_3$ is increased for the ``physical'' (R,NS) and (NS,NS) fermions.

\section{Bosonization and the Sine-Gordon Model}

We used the bosonization technique to compute the entanglement entropy of a free massless Dirac fermion. Even after turning on the mass, one can still employ the bosonization from massive Dirac fermions to the sine-Gordon model:
\begin{align}
	\CL_{SG} = \frac{1}{8\pi} \partial_\mu\phi \partial^\mu \phi + \lambda \cos \phi \ ,
\end{align}
where $\lambda$ is proportional to the mass of the Dirac fermion: $\lambda = \frac{m}{\pi\epsilon L}$ \cite{Zamolodchikov:1995xk,Lukyanov:1996jj}.
Then, the leading correction of the partition function $Z_{k,\nu}$ starts from the $O(\lambda^2)$ term due to the charge conservation of vertex operators:
\begin{align}
	Z_{k,\nu}(m) = Z_{k,\nu} + \frac{\lambda^2}{2} \int d^2 x d^2 y \,
	\langle \cos \phi(x) \cos \phi(y) \prod_{a=1}^p V_k(u_a) V_{-k}(v_a) \rangle_\nu + O(\lambda^4)\ .
\end{align}
The integrand can be evaluated by using eq.\ \eqref{VertexCorr} as follows:
\begin{align}
	&\langle \cos \phi(x) \cos \phi(y) \prod_{a=1}^p V_k(u_a) V_{-k}(v_a) \rangle_\nu 
	= \frac{1}{4}\langle\prod_{a=1}^p V_k(u_a) V_{-k}(v_a)\rangle_\nu \left[A_{k,\nu}(x,y)+ A_{k,\nu}(y,x)\right]\ ,
\end{align}
where 
\begin{align}\label{Aknu}
		A_{k,\nu}(x,y) \equiv \left| \frac{\vartheta_\nu(\frac{k\sum_a(v_a - u_a)}{n}  + x-y|\tau)}{\vartheta_\nu(\frac{k\sum_a(v_a - u_a)}{n} |\tau)}
		 \frac{\epsilon\, \partial_z \vartheta_1(0|\tau)}{\vartheta_1(y-x|\tau)} \right|^2 \prod_{a=1}^p \left| \frac{\vartheta_1(v_a - x|\tau) \vartheta_1(u_a - y|\tau)}{\vartheta_1(u_a - x|\tau) \vartheta_1(v_a - y|\tau)}\right|^{\frac{2k}{n}} \ .
\end{align}

\begin{figure}[htbp]
	\centering
	\includegraphics[width=8cm]{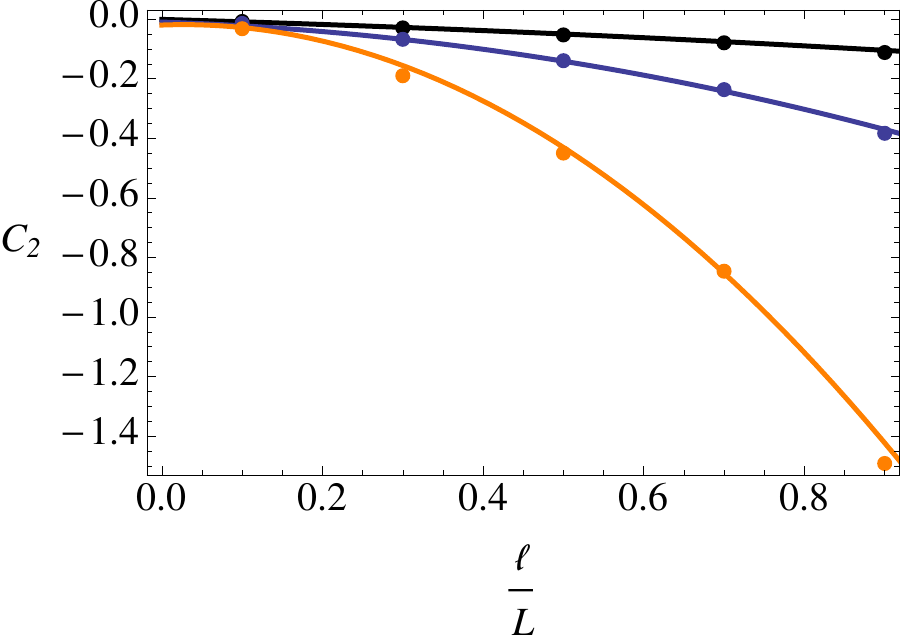}
	\caption{
	The $\ell$ dependence of the $O(m^2)$ correction to the $n=2$ R\'enyi entropy for $\nu=2$.  The curves are produced by numerical integration of (\ref{Cndef}).  The points are from a lattice computation.  From top to bottom, $\beta = 1/2$, 1, and 2.
	 }
	\label{masscoeff}
\end{figure}

At leading order, the R\'enyi entropy is given by
\begin{align}
\label{m2corr}
	S_n^{(\nu)} (m) = S^{(\nu)}_n (0) + C_n\, m^2 + O(m^4) \ ,
\end{align}
where the coefficient $C_n$ of $m^2$ is defined by
\begin{align}
\label{Cndef}
	C_n = \frac{1}{1-n} \frac{1}{4\pi^2 (\epsilon L)^2} \int d^2 x\, d^2 y\, \sum_{k=-\frac{n-1}{2}}^\frac{n-1}{2}A_{k,\nu}(x,y) \ .
\end{align}
The four dimensional integral is too complicated to evaluate analytically and we shall rely on a numerical computation after isolating and showing the trivial nature of the UV divergence.
Since $2k/n<1$, there are no poles at $x,y=u_a, v_a$ in the integrand $A_{k,\nu}(x,y)$. 
A possible divergence comes from the point $x=y$ where $\vartheta_1 (y-x|\tau) \sim y-x$.
Expanding the remainder of eq.\ \eqref{Aknu} and summing it over $k$, 
one obtains the following series around $x=y$:
\begin{align}
	\sum_{k= - \frac{n-1}{2}}^\frac{n-1}{2} \left| \frac{\vartheta_\nu(\frac{k\sum_a(v_a - u_a)}{n}  + x-y|\tau)}{\vartheta_\nu(\frac{k\sum_a(v_a - u_a)}{n} |\tau)}\right|^2 & \prod_{a=1}^p \left| \frac{\vartheta_1(v_a - x|\tau) \vartheta_1(u_a - y|\tau)}{\vartheta_1(u_a - x|\tau) \vartheta_1(v_a - y|\tau)}\right|^{\frac{2k}{n}}  = \\
	& \hspace{20mm} 1 + O((x-y)^2, (\bar x - \bar y)^2, |x-y|^2) \ . \nonumber
\end{align}
Therefore, the singular part of the integrand is 
\begin{align}
	\sum_{k= - \frac{n-1}{2}}^\frac{n-1}{2} A_{k,\nu}(x,y) =\frac{\epsilon^2}{|x-y|^2} \left[ 1 + O((x-y)^2, (\bar x - \bar y)^2, |x-y|^2) \right] \ .
\end{align}
The integration measure gives a factor of $|x-y|$ near $x\sim y$, and we end up with 
a single pole there.
This single pole gives rise to the UV divergence after integration, but the divergence is independent of the size of the intervals. Since we are interested in the physics depending on the size, we will throw the divergence away and get a finite result in the end. 
Fig.\ \ref{masscoeff} shows the result of a numerical integration of eq.\ (\ref{Cndef}) for one interval of width $v_1 - u_1 = \ell/L$.  We find good agreement with a lattice computation described in section \ref{sec:lattice}.

The IR divergence is absent on a torus, but it appears in the flat spacetime limit. 
It is worth looking into what happens in this case.
The function $A_{k,\nu}(x,y)$ reduces to the correlation function of the vertex operators on 
a flat space:\footnote{The variables $x,y,u_a,v_b$ and $\epsilon$ are dimensionless on a torus, but they have dimensions of length in the flat spacetime limit. We will use the same symbols for simplicity.}
\begin{align}\label{AknuFlat}
	A_{k,\nu}(x,y) = \frac{\epsilon^2}{|y-x|^2} \prod_{a = 1}^p \left| \frac{(v_a - x)(u_a - y)}{(u_a - x)(v_a - y)} \right|^{\frac{2k}{n}} \ .
\end{align} 
The most divergent term will come from the region $x,y \sim \Lambda$ where $\Lambda$ is the IR cut-off scale.
The expansion of $A_{k,\nu}(x,y)$ around large $x$ and $y$ is enough to compute the IR divergence:
\begin{align}
	\sum_{k= - \frac{n-1}{2}}^\frac{n-1}{2} A_{k,\nu}(x,y) = \frac{\epsilon^2}{|y-x|^2} \left[ n + \frac{n^2 - 1}{12n}  \frac{\left[ |x|^2 (y + \bar y) - |y|^2 (x+\bar x) \right]^2}{2 |x|^4 |y|^4}  \ell_t^2 + \cdots \right] \ ,
	\label{Aknuinter}
\end{align}
where $\ell_t = \sum_a(v_a - u_a)$ is the total length of the intervals.
The leading term is an $\ell_t$ independent 
 IR divergence, and we drop it below.
Performing the integral over $x$ and $y$, we obtain 
\begin{align}\label{OurIR}
	S_n(m) = S_n(0) - \frac{1+n}{12n} (m \ell_t)^2 \log^2 \Lambda + \cdots \ .
\end{align} 
This small mass expansion is strikingly similar to the result (\ref{Ssmallmsingle}) 
of refs.\ \cite{Casini:2005rm,Casini:2009vk} reviewed in the introduction.

We are working in a limit $m \ll 1/L, T$, and our IR cutoff is naively given by the size of the torus $L$ and $\beta$.  If we can commute the order of limits, we may identify the IR cutoff instead with the inverse 
mass of the Dirac fermion, $\Lambda = 1/m$, and then our result \eqref{OurIR} agrees with \eqref{Ssmallmsingle}. We will see below that the limits commute for the $\nu=3$ spin structure but not for $\nu=2$.  In the $\nu=2$ case, there is an extra contribution from $\delta s(n,r)$ in eq.\ \eqref{S2delta} that needs to be removed when the limits are exchanged.

\section{Massive Fermion on the Lattice}\label{sec:lattice}

The Hamiltonian of a Dirac fermion on a circle of radius $L$ can be derived from the corresponding Lagrangian density (\ref{LDF}):
\begin{align}
	H = \int_0^L dx\, \Psi^\dagger  (- i \gamma^0 \gamma^1 \partial_x + m \gamma^0) \Psi \ .
\end{align}
To put the fermion on the lattice, we discretize the circle into $N$ points with a lattice separation $\epsilon = L/N$
\begin{align}
	H = \sum_{j=1}^N \left[ -\frac{i}{2} (\Psi_j^\dagger \sigma^3 \Psi_{j+1} - \Psi_{j+1}^\dagger \sigma^3 \Psi_{j} ) + m\epsilon\, \Psi^\dagger_j \sigma^1 \Psi_j \right] \ .
\end{align}
The canonical anti-commutation relations are $\{ \Psi_{j,\alpha}, \Psi_{k,\beta}^\dagger\} = \delta_{jk}\, \delta_{\alpha\beta}$, and $\alpha,\beta$ are the spinor indices.
To diagonalize the Hamiltonian, we expand the Dirac field as follows:
\begin{align}\label{Psi}
	\Psi_j = \frac{1}{\sqrt{L}} \sum_{a=1-\nu_1}^{N-\nu_1} \frac{1}{\sqrt{2\omega(\theta_a)}} \left[ b(\theta_a) \, u_a\,  e^{-i \theta_a j} + d^\dagger(\theta_a)\, v_a\, e^{i \theta_a j}\right] \ .
\end{align}
To satisfy periodic ($\nu_1=0$) or antiperiodic ($\nu_1=1/2$) boundary conditions around the circle, we set $\theta_a = \frac{2\pi a}{N}$.
The energy $\omega(\theta_a)$ is defined by
\begin{align}
	\omega(\theta_a)^2 = m^2 + \frac{\sin^2 \theta_a}{\epsilon^2} \ .
\end{align}
This dispersion relation exhibits the classic doubling problem of fermions on the lattice.  Our concern with finite size effects, however, introduces an additional subtlety.  If we take $N$ even, then we get two copies of either a $\nu_1 = 0$ or a $\nu_1=1/2$ fermion.  If we take $N$ odd, then the second copy has the continuum spectrum with spatial periodicity opposite that indicated by $\nu_1$.  
Numerically, we have observed that the entropy in this case corresponds to a $\nu_1=0$ plus a $\nu_1=1/2$ fermion.
To keep things simple, we will assume $N$ is even from now on and then divide our entropies by two when comparing with the analytic results from earlier in the paper.

The $u_a$ and $v_a$ are normalized such that\footnote{
 The vectors $u_a$ and $v_a$ satisfy the discretized Dirac equations
\begin{align}\label{DiracEq}
	\left(\omega(\theta_a) \gamma^0 + \frac{\sin\theta_a}{\epsilon} \gamma^1 - m\right) u_a &= 0 \ , \nonumber\\
	\left(\omega(\theta_a) \gamma^0 + \frac{\sin\theta_a}{\epsilon} \gamma^1 + m\right) v_a &= 0 \ .
\end{align}
We demand that the $u_a$ and $v_a$ satisfy the normalization and orthogonality conditions
\begin{align}\label{normalize}
	u_a^\dagger u_a &= v_a^\dagger v_a = 2\omega(\theta_a) \ , \nonumber\\
	u_a^\dagger v_{N-a} &= v_a^\dagger u_{N-a} = 0 \ .
\end{align}
One can explicitly find the vectors satisfying \eqref{DiracEq} and \eqref{normalize} 
\begin{align*}
	u_a &= \left(\sqrt{\omega(\theta_a) - \frac{\sin \theta_a}{\epsilon}}, \sqrt{\omega(\theta_a) + \frac{\sin \theta_a}{\epsilon}} \right) \ , \\
	v_a &= \left(-\sqrt{\omega(\theta_a) - \frac{\sin \theta_a}{\epsilon}}, \sqrt{\omega(\theta_a) +\frac{\sin \theta_a}{\epsilon}} \right) \ .
\end{align*}
}
$\{ b(\theta_a), b^\dagger (\theta_b) \} = \delta_{ab}$ and $\{ d(\theta_a), d^\dagger (\theta_b) \} = \delta_{ab}$. 
The Hamiltonian is diagonalized as\footnote{We remove the infinite constant coming from the commutation of $d$ and $d^\dagger$. In other words, we fill out the Dirac sea.}
\begin{align}
	H  = \sum_{a=1-\nu_1}^{N-\nu_1}\omega(\theta_a) \left[ b^\dagger (\theta_a) b(\theta_a) + d^\dagger(\theta_a) d(\theta_a) \right] \ .
\end{align}
In the lattice model, the fermion number operator is given by
\begin{align}
	F = \sum_{a = 1-\nu_1}^{N-\nu_1} \left(b^\dagger(\theta_a) b(\theta_a) - d^\dagger(\theta_a) d(\theta_a) \right) \ .
\end{align}

Introducing a chemical potential $\mu$ conjugate to $F$, the density matrix can be written in terms of $H$ in the standard way:
\begin{equation}\label{TDM}
\rho = \frac{(-1)^{(1-2 \nu_2)F} e^{-(H + \mu F) /T}}{\tr[(-1)^{(1-2 \nu_2)F}  e^{-(H+\mu F)/T}]} \ .
\end{equation}
We have introduced a factor of $(-1)^F$ to allow for spin structures periodic in the time direction.
Expectation values are defined as $\langle X \rangle \equiv \tr (\rho X)$.  
A short calculation yields the two-point correlation function of two $\Psi$ fields:
\begin{align}
	\langle \Psi_j \Psi_k^\dagger \rangle = &\frac{1}{2L} \sum_{a=1-\nu_1}^{N-\nu_1} e^{i \theta_a (j-k)} \Bigg[ 
		\left(1 + \frac{\sinh (\beta \mu)}{\cosh (\beta\mu) + (-1)^{2\nu_2 +1} \cosh(\beta \omega(\theta_a))} \right) \nonumber \\
		&\qquad\qquad + \left( \frac{\sin \theta_a}{\omega (\theta_a) \epsilon} \sigma_3 + \frac{m}{\omega(\theta_a)} \sigma_1 \right) \frac{\sinh(\beta\omega(\theta_a))}{ (-1)^{2\nu_2 +1}\cosh(\beta\mu) + \cosh(\beta\omega(\theta_a))} 
	\Bigg] \ .
\end{align}
Note that the argument in section \ref{sec:ACP} implies the $\nu_2=0$  sector is obtained from the $\nu_2 = 1/2$ sector
by shifting $\mu \to \mu - i \pi /\beta$.  The form of the two-point function is consistent with this observation.

It is possible to calculate R\'enyi entropies from the matrix $C^{(\nu)} = \epsilon \langle \Psi \Psi^\dagger \rangle$.  
Consider a region $A$, which may consist of many disjoint subintervals of the circle, and the corresponding reduced density matrix $\rho_A$.  
We restrict $C^{(\nu)}_{jk}$ such that $j$ and $k$ run only over sites in $A$.  Call the restricted two-point function $C^{(\nu)}_A$. 
Remarkably, for a free spinor field,
the reduced density matrix $\rho_A \sim e^{-H_A}$ 
can be written in terms of a free particle Hamiltonian $H_A = \sum_k \epsilon_k b_k^\dagger b_k$ (see for example \cite{Casini:2009sr, EislerPeschel}).  
Moreover, there is a one-to-one correspondence between eigenvalues $\lambda_j$ of $C_A$ and the energies $\epsilon_j$:
\begin{equation}
\label{lambdaepsilonrel}
\lambda_j = \frac{1}{1 - (-1)^{2\nu_2} e^{\epsilon_j}} \ .
\end{equation}
Given this relation, it is a short exercise to demonstrate that the R\'enyi entropies are
\begin{align}\label{LatticeRE}
	S^{(\nu)}_n = \frac{1}{1-n}\tr  \log \left[  (1-C^{(\nu)}_A)^n +  (C^{(\nu)}_A)^n \right] \ ,
\end{align}
where $C^{(\nu)}_A$ is the restricted two-point function. 

Before proceeding, we make two quick observations about the eigenvalue distribution of $C^{(\nu)}_A$. 
From the trace of $C^{(\nu)}_A$, we see that when the chemical potential vanishes, 
$\sum_j \lambda_j = n$ where $n$ is the total length of $A$.
Next, from the relation (\ref{lambdaepsilonrel}), it is clear that in the thermal case ($\nu_2=1/2$), $\lambda_j$ is bounded between zero and one.  
Provided $0 \leq \lambda_j \leq 1$, we can take a sensible $n \to 1$ limit of (\ref{LatticeRE}) and derive the entanglement entropy 
\begin{align}\label{LatticeEE}
	S^{(\nu)} = - \tr \left[ (1-C^{(\nu)}_A) \log (1-C^{(\nu)}_A) + C^{(\nu)}_A \log C^{(\nu)}_A \right] \ .
\end{align}

\subsection{Comparison to the analytic results}
\begin{figure}[htbp]
	\centering
	\includegraphics[width=5cm]{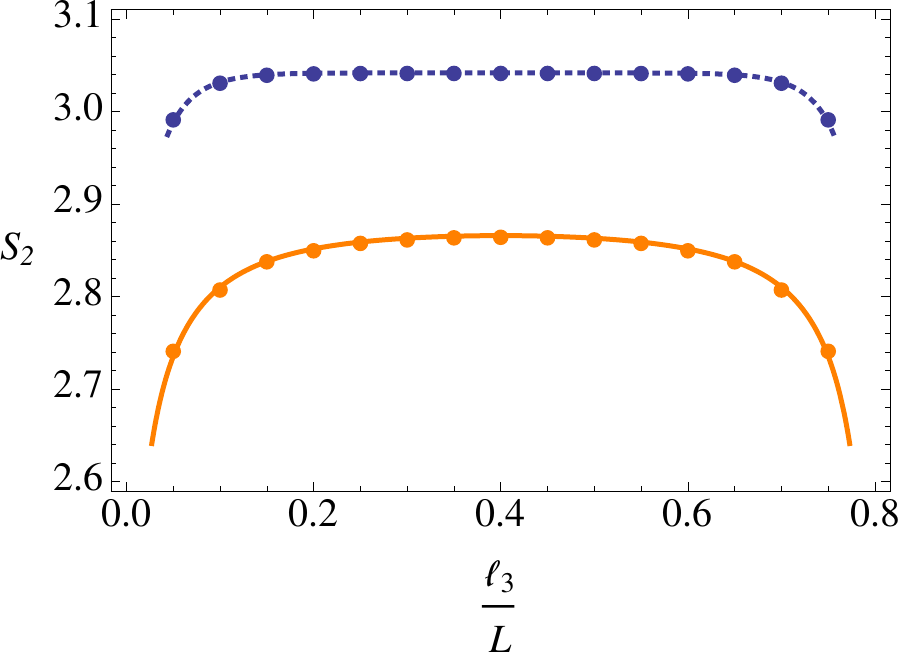}\quad
	\includegraphics[width=5cm]{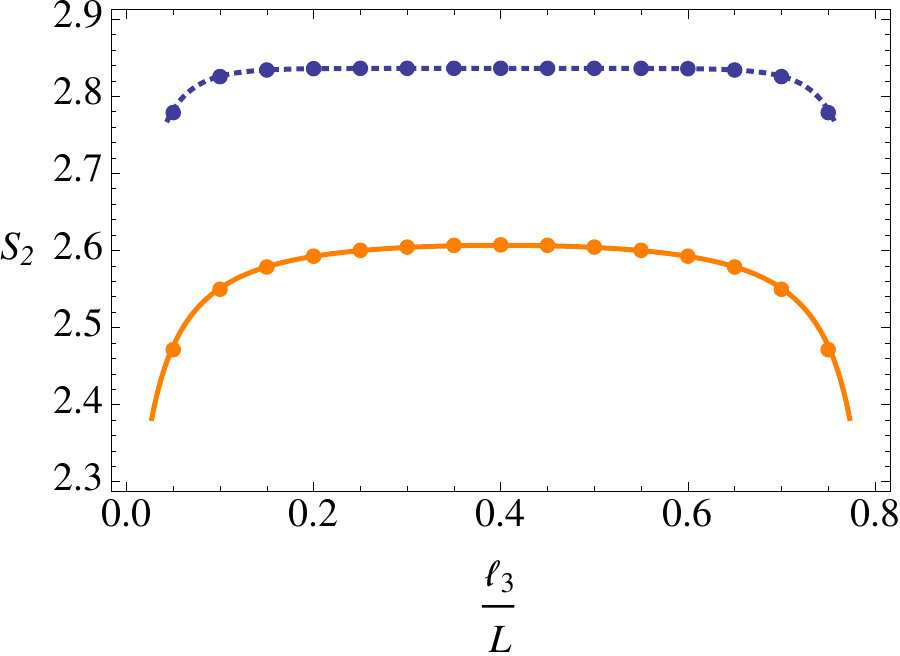}\quad
	\includegraphics[width=5cm]{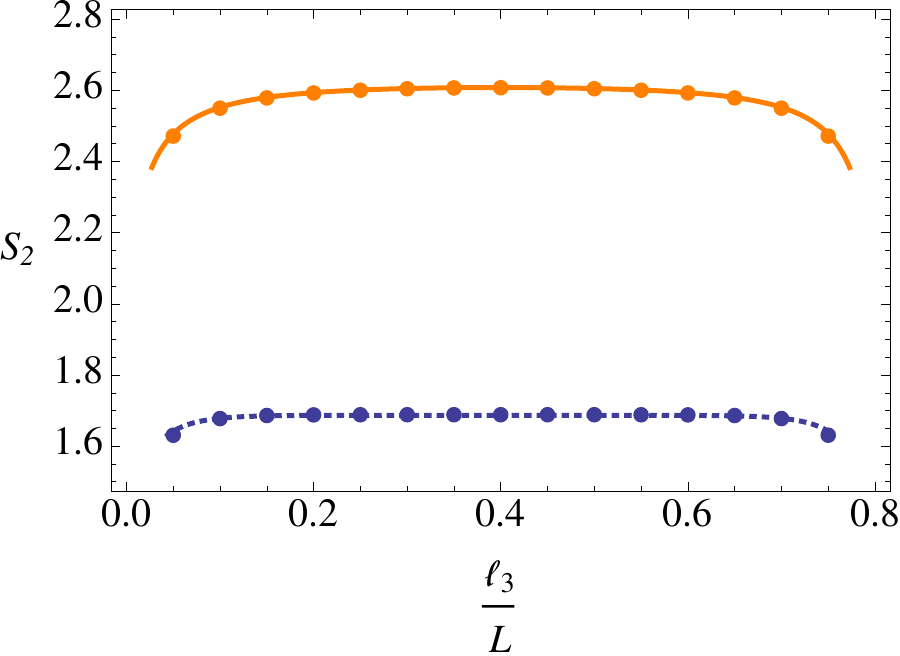}
	\caption{The R\'enyi entropy for $n=2$ of two intervals of width $\ell_1 = \ell_2=L/10$ whose distance is $\ell_3$. The $\nu=2$ [Left], $\nu=3$ [Middle] and $\nu=4$ [Right] sectors are depicted. The curves are analytic and the dots are numerical. The blue dotted and orange solid curves are for $\beta=1/5$ and $10$, respectively.}
	\label{TwoIntervalS2L3}
\end{figure}

\begin{figure}[htbp]
	\centering
	\includegraphics[width=5cm]{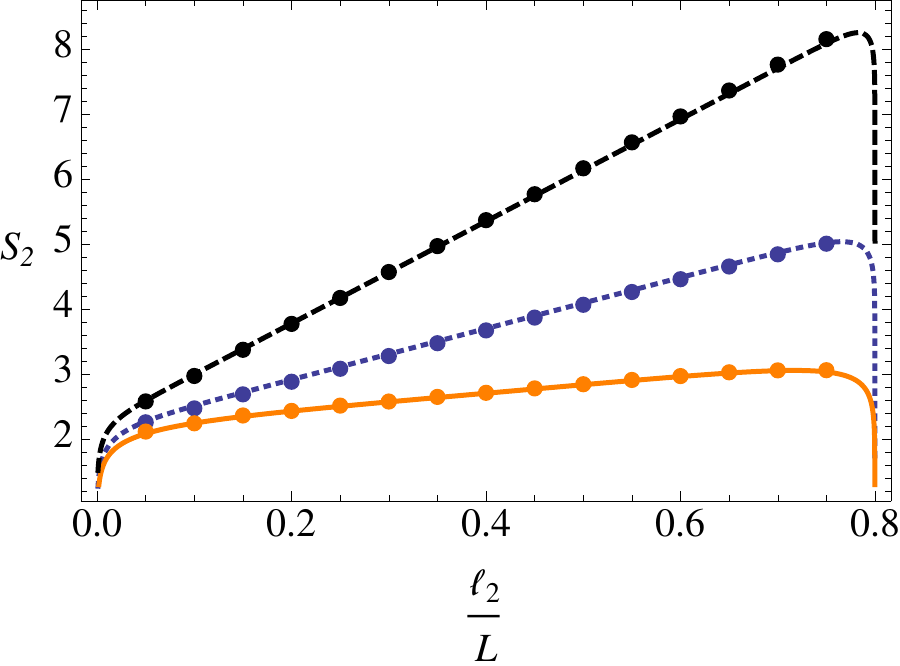}\quad
	\includegraphics[width=5cm]{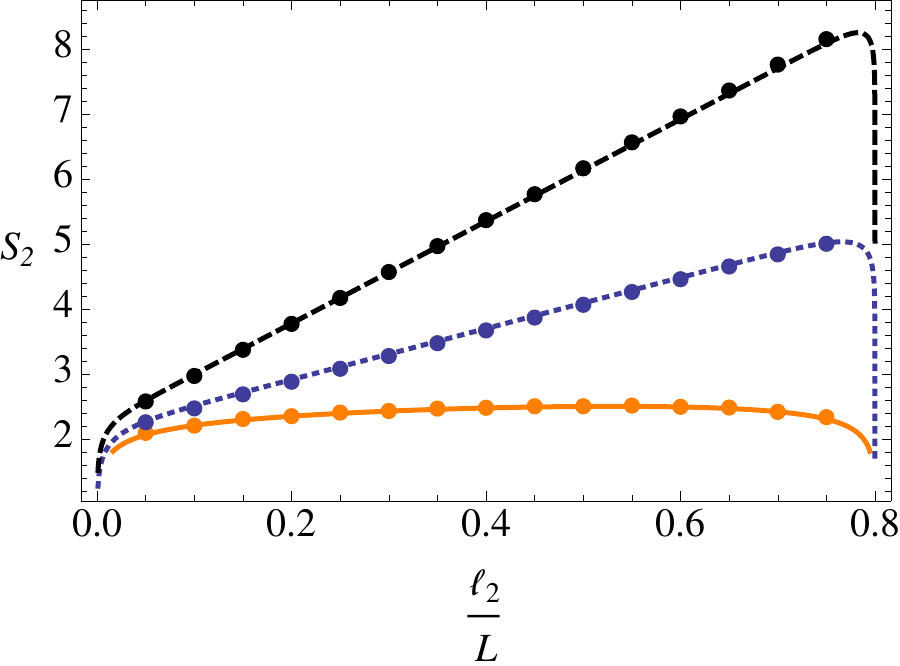}\quad
	\includegraphics[width=5cm]{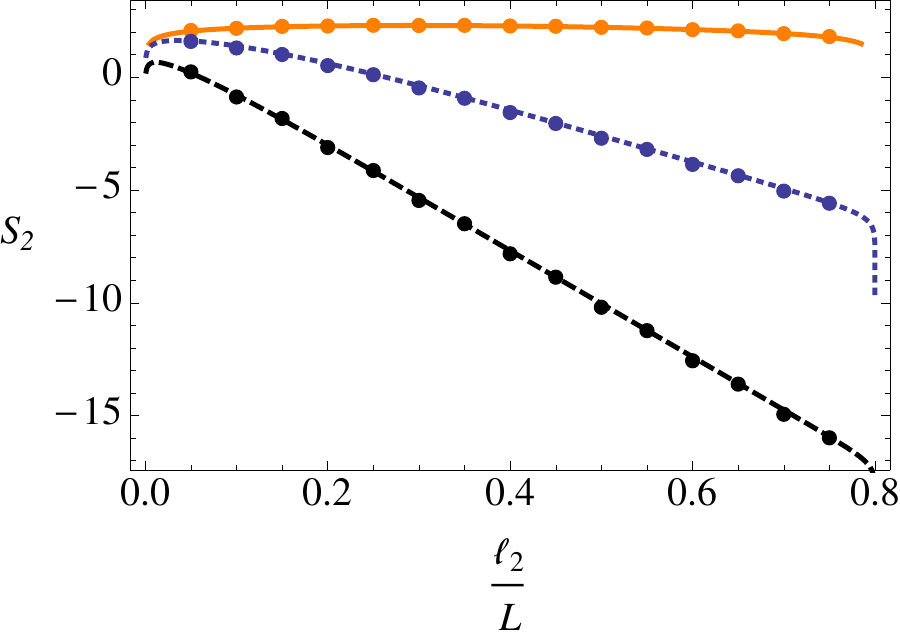}
	\caption{The R\'enyi entropy for $n=2$ of two intervals of width $\ell_1=L/10$ and $\ell_2$. The distance between the intervals is fixed to $\ell_3=L/10$ and $\ell_2$ is varied. The $\nu=2$ [Left], $\nu=3$ [Middle] and $\nu=4$ [Right] sectors are depicted. The curves are analytic and the dots are numerical. The orange solid, blue dotted and black dashed curves are for $\beta=1/10,1/5$ and $1$.}
	\label{TwoIntervalS2L2}
\end{figure}

In section \ref{sec:Massless}, using bosonization, 
we obtained analytic formulae for the R\'enyi entropies of a massless Dirac fermion.  
To gain confidence in our methods, we  compare the lattice calculation for a massless fermion with these analytic formulae.
Consider the R\'enyi entropy for $n=2$ of two intervals of width $\ell_1$ and $\ell_2$ separated by a distance of $\ell_3$.
In Fig.\,\ref{TwoIntervalS2L3}, we plot the entropies in the $\nu=2,3,4$ sectors by changing the distance $\ell_3$ with fixed $\ell_1 = \ell_2 = L/10$.
The blue (dotted) and orange (solid) curves are the analytic results for $\beta=1/5$ and $10$, respectively.
The dots are plotted using the lattice computation, which nicely agree with the analytic curves up to 
a constant. Since the R\'enyi entropy is always UV divergent, a constant shift is allowed to match the analytic and numerical results.
The other case is studied in Fig.\,\ref{TwoIntervalS2L2} by varying $\ell_2$ with fixed $\ell_1=\ell_3 = L/10$.
The analytic and numerical results perfectly fit each other for various temperatures.

\subsection{Small mass vs. small temperature}
An interesting feature of the (R,NS) fermion is that the $T\to 0$ and $m \to 0$ limits do not commute.  
In the zero mass and zero chemical potential limit, the two point function becomes
\begin{equation}
\label{mzeroC}
\lim_{m \to 0} C^{(\nu)}_{jk}= \frac{1}{2N} \sum_{a=1-\nu_1}^{N-\nu_1} e^{i \theta_a(j-k)} \left[ 1 + \sigma_3 \sgn (\sin \theta_a) 
\left( \tanh \frac{\omega(\theta_a)}{2T} \right)^{4 \nu_2 - 1} \right]\ .
\end{equation}
If we further take the zero $T$ limit of (\ref{mzeroC}) for the (R,NS) and (NS,NS) fermions, we obtain respectively
\begin{align}
\lim_{T \to 0} \lim_{m \to 0} C_{jk}^{(2)} &=
\frac{1}{2} \delta_{jk} + (1- \delta_{jk}) \frac{i \sigma_3}{N} \left| \sin \frac{\pi(j-k)}{2} \right| \cot \frac{\pi(j-k)}{N} \ , \\
\lim_{T \to 0} \lim_{m \to 0} C_{jk}^{(3)} &=
\frac{1}{2} \delta_{jk} +  (1- \delta_{jk}) \frac{i \sigma_3}{N} \left| \sin \frac{\pi(j-k)}{2} \right| \csc \frac{\pi(j-k)}{N} \ .
\end{align}
Because the (NS,NS) theory is gapped even for $m=0$, we find
\[
\lim_{T \to 0} \lim_{m \to 0} C_{jk}^{(3)} = \lim_{m \to 0} \lim_{T \to 0} C_{jk}^{(3)} \ .
\]
However, in the (R,NS) case, we find instead that
\begin{equation}
\lim_{m \to 0} \lim_{T \to 0} C_{jk}^{(2)} =
\lim_{T \to 0} \lim_{m \to 0} C_{jk}^{(2)}  
+ \frac{\sigma_1}{N} \left| \cos \frac{\pi(j-k)}{2} \right| \ .
\end{equation}

\begin{figure}
	\centering
	\includegraphics[width=7cm]{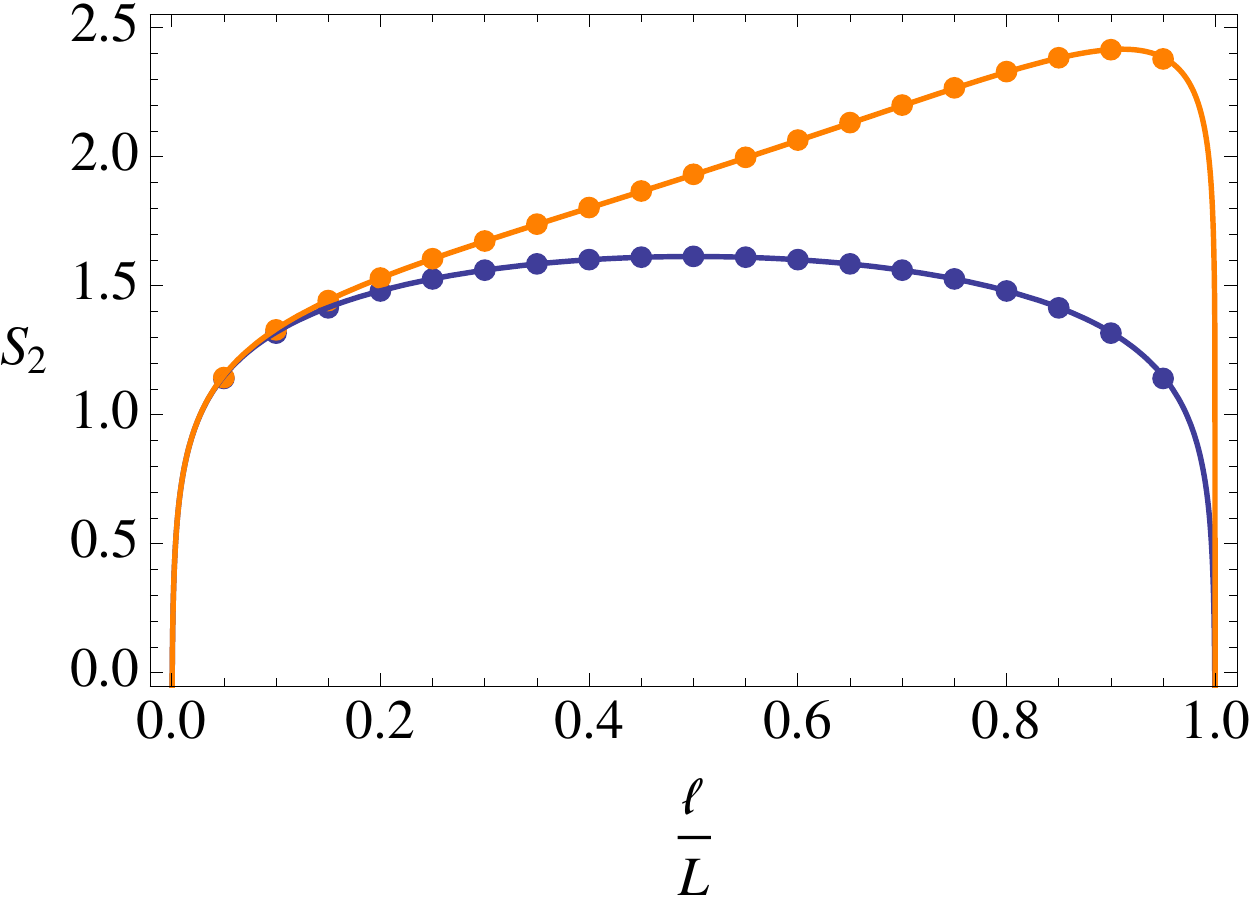}
	\quad
	\includegraphics[width=7cm]{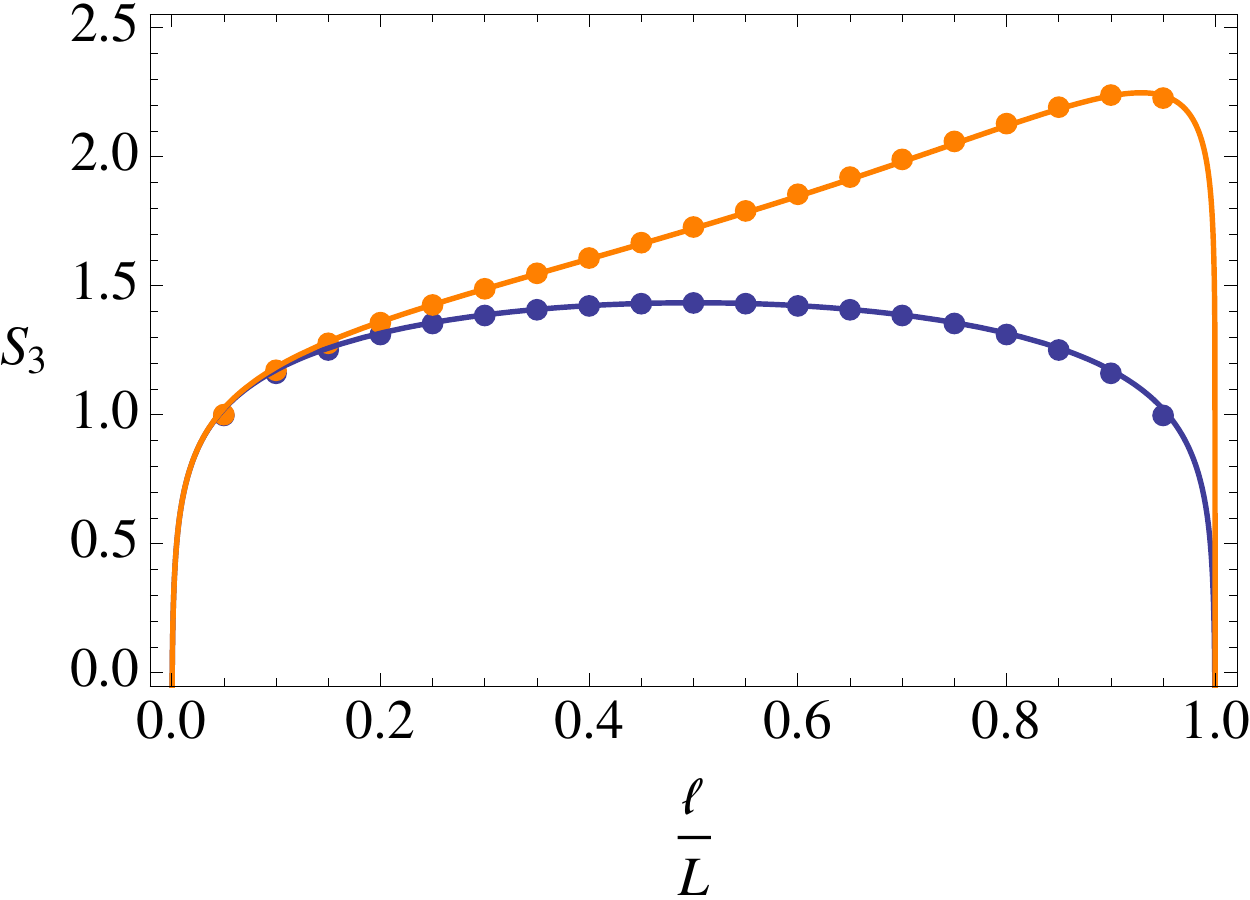}
	\caption{The single interval R\'enyi entropy for $\nu =2$: $n=2$ [Left] and $n=3$ [Right]. 
	 In both cases, the $\lim_{T\to 0} \lim_{m \to 0}$ curve (orange) lies above and the $\lim_{m\to 0} \lim_{T \to 0}$ curve (blue) 
	 lies below.
	 	 The points were computed using the lattice.
	 }
	\label{noncom}
\end{figure}

Let us restrict to the case where $A$ is a single interval of length $n$.  
It turns out that $\lim_{m \to 0} \lim_{T \to 0} C_{jk}^{(2)} \equiv C_R$ and $\lim_{T \to 0} \lim_{m \to 0} C_{jk}^{(3)} \equiv C_{NS}$ have the same eigenvalue spectrum, provided $n$ is even.  There is a similarity transform which relates the two
\begin{equation}
C_{NS} \cdot M = M \cdot C_R \ ,
\end{equation}
where
\begin{equation}
M_{jk} = \delta_{jk} \cos \frac{(j-1)\pi}{N} - \delta_{n+1-j,k} \, \sigma_2 \sin \frac{(j-1)\pi}{N} \ ,
\end{equation}
$1 \leq j,k \leq n$.
(For odd $n$, the eigenvalue spectra must then approach each other in the large $N$ limit by continuity.)
This equivalence means 
we can compute
$
\left[ \lim_{T \to 0}, \lim_{m \to 0} \right] S(A)
$
for the (R,NS) fermion
using our bosonization results:
\begin{align}
\left[ \lim_{T \to 0}, \lim_{m \to 0} \right] S^{(2)}_n(m,T)
&=
\lim_{T \to 0} (S_n^{(2)}(0,T) - S_n^{(3)}(0,T)) 
\nonumber \\
&=
\lim_{T \to 0} (S_{n,1}^{(2)} - S_{n,1}^{(3)}) 
\nonumber \\
&=
\delta s(n,r) 
\label{deltasresult} \ . 
\end{align}
For the entanglement entropy, we find that $\left[ \lim_{T \to 0}, \lim_{m \to 0} \right] S^{(2)}(m,T) = \delta s(1,r)$.
The non-commuting nature of these limits is shown in Fig.~\ref{noncom}.
Numerics suggest that the result (\ref{deltasresult}) holds for multiple intervals as well.

\subsection{Small mass and temperature}

For theories with a mass gap $m_{\rm gap}$, ref.\ \cite{Herzog:2012bw} conjectured that the temperature dependent portion of the entanglement entropy should have an exponential scaling dependence $e^{-m_{\rm gap} /T}$ in the range $m_{\rm gap} \gg T$.  
More precisely, given an interval $A$ and its complement $B$, the conjecture posits that 
\[
S_A(T) - S_B(T) \sim S_A(T) - S_A(0) \sim e^{-m_{\rm gap} /T} \ . 
\]
In this section, we provide further evidence for this conjecture.

For the $\nu_1=1/2$ fermions, the ground state is gapped with $m_{\rm gap} = \pi / L$.  
Reassuringly, our low temperature expansions (\ref{S113}) and (\ref{S114L}) for $\lim_{n\to 1} S_{n,1}^{(3)}$ 
and $\lim_{n\to 1} S_{n,1}^{(4)}$ yield precisely such scaling behavior, and we get the prefactor:
\begin{align}
S_A(T) - S_A(0) &=  \pm 4 (1 - \pi r \cot (\pi  r) ) e^{-\pi/L T} + O(e^{-2 \pi/ L T}) \ , \\
S_A(T) - S_B(T) &= \mp 4 \pi  \cot(\pi r) e^{-\pi / L T} + O(e^{-2 \pi/ L T}) \ ,
\end{align}
where 
the top choice of sign corresponds to $\nu=3$ and the bottom to $\nu=4$. 
The region $A$ is taken to have size $r L$.  Similar scaling behavior holds for the R\'enyi entropies and can be computed from eqs.~(\ref{Sn13}) and (\ref{Sn14}).

\begin{figure}[htbp]
	\centering
	\includegraphics[width=7cm]{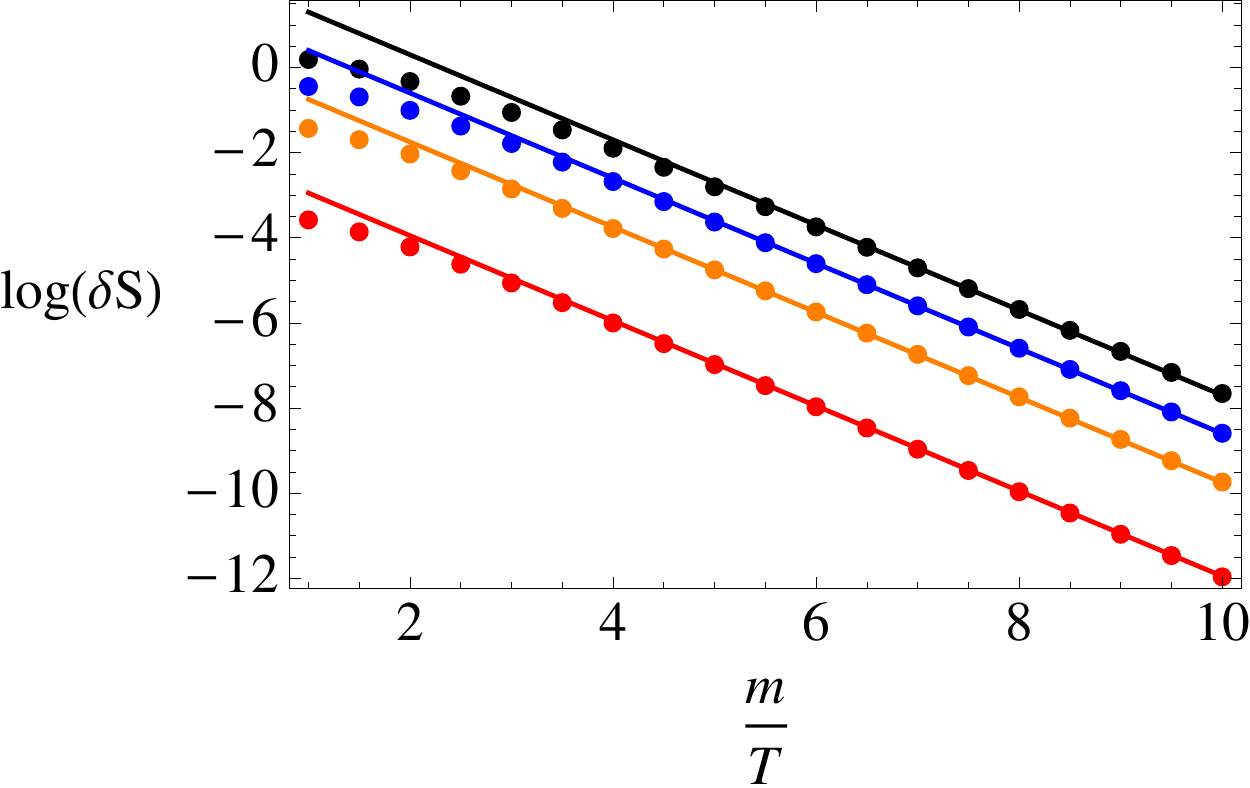}
	\quad
	\includegraphics[width=7cm]{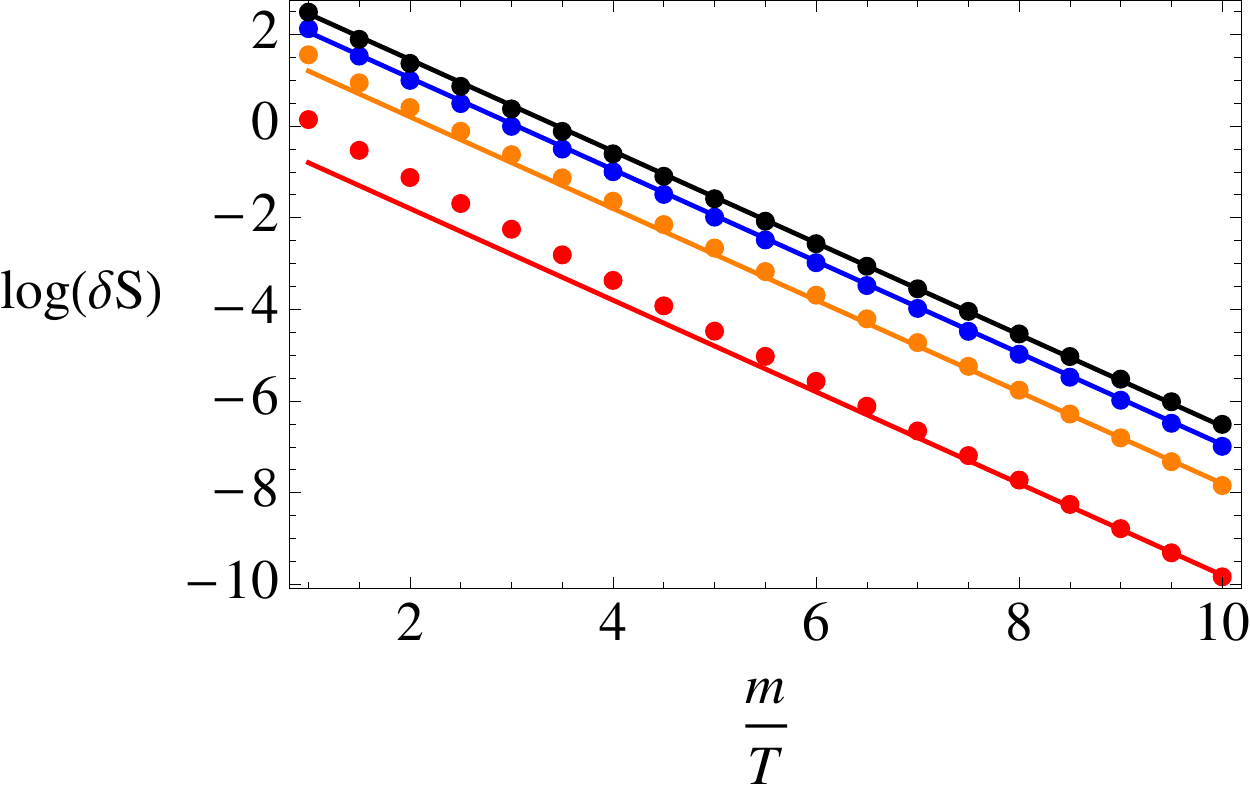}
	\caption{The entanglement entropy difference $\delta S = S(T) - S(0)$ for (R,NS) fermions: [Left] $mL = 1/10$; [Right] $mL = 10$.  The points are computed from a lattice, and the lines are fits with slope -1.  
	From bottom to top, $\ell / L = 1/10$, 3/10, 1/2, 7/10.}
	\label{expdep}
\end{figure}

We also investigate the scaling behavior for spatially periodic $\nu_1 = 0$ fermions where we introduce an $m \neq 0$ by hand.  In this case, we have no analytic results to offer, but we can use the lattice to calculate the entanglement entropy numerically.  
We compute $\delta S = S(T) - S(0)$ for the (R,NS) fermion and a single interval.  Fig.\ \ref{expdep} clearly shows $e^{-m/T}$ scaling in the region $m \gg T$, both for small mass $mL = 1/10$ and large mass $mL = 10$.

\section{Discussion}
Our bosonization 
method of computing the R\'enyi entropy for a massive Dirac fermion is perturbative in the mass, and we would like to do better. 
As reviewed in the introduction, in flat spacetime, ref.\ \cite{Casini:2005rm} obtained a non-perturbative relation between the single interval R\'enyi entropy 
and a solution to the Painlev\'e V equation.  This non-perturbative relation uses a result of ref.\ \cite{Bernard:1994re}
for the sine-Gordon model.
Similar arguments may be useful for investigating the behavior of the massive fermion on a torus.

Another possible non-perturbative approach is to use the lattice. The two fermion correlation function
matrix $C^{(\nu)}_A$ that we derived above is Toeplitz.
The R\'enyi entropy can be expressed in terms of a contour integral over the characteristic polynomial of $C^{(\nu)}_A$.
Mathematical techniques such as the Szeg\"o limit theorem and further generalizations such as the Fisher-Hartwig formula are available for taking such determinants.
Indeed, such techniques have already been used to study the XY model \cite{jin2004quantum,Its2009fisher}. 
Through a Jordan-Wigner transformation, the continuum limit of the XY model can be related to nonrelativistic free fermions.

A field theory with a mass gap can be implemented geometrically by putting a gauge theory on a compact space.
Such field theories sometimes have holographic duals with AdS geometries where the compact space is the conformal boundary \cite{Witten:1998zw}.  Several authors have studied holographic entanglement entropies in these backgrounds \cite{Nishioka:2006gr,Klebanov:2007ws,Pakman:2008ui}.
In the case of the mutual information, 
for strip like regions, there is a ``phase transition'' where the entanglement entropy is nonzero for two strips close together but vanishes once the strips become sufficiently far apart.  In our case, we do not expect to have a phase transition given that we have a finite number of degrees of freedom and work in finite volume.  Nevertheless, we do see that the mutual information is exponentially suppressed for large separations and high temperatures (at least for the ``physical'' $\nu=2$ and 3 spin sectors).
In the case of temperature dependence of the entanglement entropy, 
holographic examples typically predict that quantities such as $S_A(T) - S_A(0)$ and $S_A(T) - S_{\bar A}(T)$ vanish exactly when $T \ll m_{\rm gap}$.  In our case, we again see instead exponential suppression.
Holographic theories are supposed to describe strongly coupled large-$N$ field theories and the large-$N$ effect can drive the system to a phase transition. 
Presumably, we would need $1/N$ corrections to see holographically the exponential behavior observed in this paper.
Perhaps these $1/N$ corrections could be studied by introducing higher derivative corrections, additional saddle-points in the path integral, or non-perturbative objects such as D-branes and orientifold planes.

Another interesting direction for future study is to introduce interactions between the fermions.  It is well known \cite{Coleman:1974bu,Mandelstam:1975hb} that the sine-Gordon model, for more general choice of the interaction parameter $\lambda$, fermionizes to the Thirring model which has a quartic interaction term.  On the one hand, such a quartic interaction is not compatible with the replica trick where we replaced a single fermion field on the $n$-covering space with $n$ decoupled fields living on a single torus.  On the other, one could certainly use bosonization to treat the quartic interaction perturbatively.

\subsection*{Acknowledgements}
We are grateful S.\,Abanov, J.\,Maciejko, B.\,McCoy, I.\,Klebanov, V.\,Korepin, B.\,Safdi, W.\,Siegel, M.\,Spillane, T.\,Takayanagi, T.\,Ugajin, H.\,Verlinde, and K.\,Yonekura for valuable discussions.
This work was supported in part by the National Science Foundation under Grants No.\ PHY-0844827 and PHY-0756966.  
C.~H.\ thanks the Sloan Foundation for partial support.


\appendix
\section{Theta Function Identities}
\label{app:theta}
\begin{align}
\vartheta_1(z | \tau) &=
2 e^{\pi i \tau/4} \sin (\pi z) \prod_{m=1}^\infty (1-q^m) (1- yq^m) (1-y^{-1} q^m)  \ ,\\
\vartheta_2(z | \tau) &=
2 e^{\pi i \tau/4} \cos (\pi z) \prod_{m=1}^\infty (1-q^m) (1+ yq^m) (1+y^{-1} q^m)  \ ,\\
\vartheta_3(z | \tau) &=
\prod_{m=1}^\infty  (1-q^m) (1+y q^{m-1/2})(1+y^{-1} q^{m-1/2})  \ , \\
\vartheta_4(z | \tau) &=
\prod_{m=1}^\infty (1-q^m) (1 - y q^{m-1/2})(1-y^{-1} q^{m-1/2}) \ ,
\end{align}
where $y = e^{2 \pi i z}$ and $q = e^{2 \pi i \tau}$.
We also have the S-duality relations
\begin{align}
\vartheta_1(z | \tau) &= -(-i \tau)^{-1/2} e^{-\pi i z^2 / \tau} \vartheta_1(z/\tau | -1/\tau) \ , \\
\vartheta_2(z | \tau) &= (-i \tau)^{-1/2} e^{-\pi i z^2 / \tau} \vartheta_4(z/\tau | -1/\tau) \ , \\
\vartheta_3(z | \tau) &= (-i \tau)^{-1/2} e^{-\pi i z^2 / \tau} \vartheta_3(z/\tau | -1/\tau) \ , \\
\vartheta_4(z | \tau) &= (-i \tau)^{-1/2} e^{-\pi i z^2 / \tau} \vartheta_2(z/\tau | -1/\tau) \ .
\end{align}
The periodicities of the elliptic theta functions yield 
\begin{align}\label{ShiftTheta}
\vartheta_2(z | \tau) &= -\vartheta_1(z - 1/2 | \tau) \ , \\
\vartheta_3(z | \tau) &= - y^{-1/2} q^{1/8}\vartheta_1(z - 1/2 - \tau / 2 | \tau) \ , \\
\vartheta_4(z | \tau) &= i y^{-1/2} q^{1/8}\vartheta_1(z - \tau /2 | \tau) \ .
\end{align}

\section{Time Periodic Spin Structures}
\label{app:nonthermal}
In the $\nu=1$ sector, we regulate $S_{n,1}^{(1)}$ by introducing a small chemical potential $\mu \ll 1/\beta$.
The large $\beta$ expansions for $\nu=1$ and 4 are
\begin{align}
S_{n,1}^{(1)} &= \frac{2}{1-n}  \sum_{ {{k = -\frac{n-1}{2}}\atop{k \neq 0}}}^{\frac{n-1}{2}
} \log \left| \sin \frac{ \pi k r}{n} \right| 
+ 2 \log \left| \frac{ \mu \beta}{2} \right| 
  \nonumber \\
& \qquad -\frac{4}{1-n} \sum_{j=1}^\infty \frac{1}{j} \frac{1}{e^{2 \pi  \beta j} - 1} \left( \frac{ \sin ( \pi j r)}{\sin \left( \frac{\pi j r}{n} \right) } - n \right) \ , 
\\
\label{Sn14}
S_{n,1}^{(4)} &= -\frac{2}{1-n} \sum_{j=1}^\infty \frac{1}{j\sinh \pi \beta j} \left( \frac{ \sin(\pi j r)}{\sin \left( \frac{\pi j r}{n} \right)} - n \right)  \ , \\
\lim_{n \to 1} S_{n,1}^{(1)} &= \lim_{n\to 1}  \frac{2}{1-n}  \sum_{ {{k = -\frac{n-1}{2}}\atop{k \neq 0}}}^{\frac{n-1}{2}} \log \left| \sin \frac{ \pi k r}{n} \right| 
+ 2 \log \left| \frac{ \mu \beta}{2} \right| 
\nonumber  \\
& \qquad -4 \sum_{j=1}^\infty \frac{1}{j} \frac{1}{e^{2 \pi \beta j} - 1} ( 1 - \pi j r \cot (\pi j r)) \ , 
\\
\lim_{n\to 1} S_{n,1}^{(4)} &= -2 \sum_{j=1}^\infty \frac{1}{j } \frac{1 - \pi j r \cot (\pi j r)}{\sinh \pi \beta j}  \label{S114L}\ .
\end{align}

The small $\beta$ expansions for $\nu=1$ and 4 are
\begin{align}
S_{n,1}^{(1)} &= \frac{1+n}{6n} \frac{\pi r^2}{ \beta}+
\frac{2}{1-n}\Biggl[ \sum_{{{k = -\frac{n-1}{2}}\atop{k \neq 0}}}^{\frac{n-1}{2}} \log \left| \sinh \frac{ \pi k r}{n\beta} \right| + \nonumber \\
& 
\hspace{5mm} 
-2 \sum_{j=1}^\infty \frac{1}{j} \frac{1}{e^{2 \pi j/\beta} - 1} \left( \frac{ \sinh \left( \frac{\pi j r}{\beta} \right)}{\sinh \left( \frac{\pi j r}{n\beta} \right) } - n \right) \Biggr]
+ 2 \log \left| \sin \frac{\mu}{2} \right| 
 \ ,
\\
\lim_{n\to 1} S_{n,1}^{(1)} &= 
\frac{\pi r^2}{3 \beta} - 4 \sum_{j=1}^\infty \frac{1}{j} \frac{1}{e^{2 \pi j/\beta} - 1} \left( 1 - \frac{\pi j r}{\beta} \coth \left( \frac{\pi j r}{\beta} \right) \right) \nonumber \\
& \hspace{5mm}
+\lim_{n\to 1} \frac{2}{1-n}\sum_{{{k = -\frac{n-1}{2}}\atop{k \neq 0}}}^{\frac{n-1}{2}} \log \left| \sinh \frac{ \pi k r}{n\beta} \right| 
+ 2 \log \left| \sin \frac{\mu}{2} \right| 
\ ,
\end{align}
\begin{align}
S_{n,1}^{(4)} &=
\frac{1+n}{6n} \frac{\pi r^2}{ \beta}+
\frac{2}{1-n}\Biggl[ \sum_{k = -\frac{n-1}{2}}^{\frac{n-1}{2}} \log \left| \cosh \frac{ \pi k r}{n\beta} \right|  \nonumber \\
&
\hspace{20mm} 
+2 \sum_{j=1}^\infty \frac{(-1)^{j+1}}{j} \frac{1}{e^{2 \pi j/\beta} - 1} \left( \frac{ \sinh \left( \frac{\pi j r}{\beta} \right)}{\sinh \left( \frac{\pi j r}{n\beta} \right) } - n \right) \Biggr] \ ,
\label{Sn14S}
\\
\label{S114}
\lim_{n\to 1} S_{n,1}^{(4)} &=
\frac{\pi r^2}{3 \beta} + 4 \sum_{j=1}^\infty \frac{(-1)^{j+1}}{j} \frac{1}{e^{2 \pi j/\beta} - 1} \left( 1 - \frac{\pi j r}{\beta} \coth \left( \frac{\pi j r}{\beta} \right) \right) \nonumber \\
& \hspace{20mm}
+\lim_{n\to 1} \frac{2}{1-n}\sum_{k = -\frac{n-1}{2}}^{\frac{n-1}{2}} \log \left| \cosh \frac{ \pi k r}{n\beta} \right|
 \ .
\end{align}

\bibliographystyle{JHEP}
\bibliography{MassiveFermion}

\providecommand{\href}[2]{#2}\begingroup\raggedright\begin{thebibliography}{10}

\bibitem{Casini:2006es}
H.~Casini and M.~Huerta, {\it {A C-Theorem for the Entanglement Entropy}},
  {\em J.Phys.} {\bf A40} (2007) 7031--7036,
  [\href{http://xxx.lanl.gov/abs/cond-mat/0610375}{{\tt cond-mat/0610375}}].

\bibitem{Casini:2012ei}
H.~Casini and M.~Huerta, {\it {On the RG Running of the Entanglement Entropy of
  a Circle}},  \href{http://xxx.lanl.gov/abs/1202.5650}{{\tt arXiv:1202.5650}}.

\bibitem{Ryu:2006bv}
S.~Ryu and T.~Takayanagi, {\it {Holographic derivation of entanglement entropy
  from AdS/CFT}},  {\em Phys.Rev.Lett.} {\bf 96} (2006) 181602,
  [\href{http://xxx.lanl.gov/abs/hep-th/0603001}{{\tt hep-th/0603001}}].

\bibitem{Ryu:2006ef}
S.~Ryu and T.~Takayanagi, {\it {Aspects of Holographic Entanglement Entropy}},
  {\em JHEP} {\bf 08} (2006) 045,
  [\href{http://xxx.lanl.gov/abs/hep-th/0605073}{{\tt hep-th/0605073}}].

\bibitem{Maldacena:1997re}
J.~M. Maldacena, {\it {The Large $N$ Limit of Superconformal Field Theories and
  Supergravity}},  {\em Adv. Theor. Math. Phys.} {\bf 2} (1998) 231--252,
  [\href{http://xxx.lanl.gov/abs/hep-th/9711200}{{\tt hep-th/9711200}}].

\bibitem{Gubser:1998bc}
S.~S. Gubser, I.~R. Klebanov, and A.~M. Polyakov, {\it {Gauge Theory
  Correlators from Non-Critical String Theory}},  {\em Phys. Lett.} {\bf B428}
  (1998) 105--114, [\href{http://xxx.lanl.gov/abs/hep-th/9802109}{{\tt
  hep-th/9802109}}].

\bibitem{Witten:1998qj}
E.~Witten, {\it {Anti-de~Sitter Space and Holography}},  {\em Adv. Theor. Math.
  Phys.} {\bf 2} (1998) 253--291,
  [\href{http://xxx.lanl.gov/abs/hep-th/9802150}{{\tt hep-th/9802150}}].

\bibitem{Calabrese:2009qy}
P.~Calabrese and J.~Cardy, {\it {Entanglement Entropy and Conformal Field
  Theory}},  {\em J. Phys.} {\bf A42} (2009) 504005,
  [\href{http://xxx.lanl.gov/abs/0905.4013}{{\tt arXiv:0905.4013}}].

\bibitem{EislerPeschel}
I.~Peschel and V.~Eisler, {\it {Reduced density matrices and entanglement
  entropy in free lattice models}},  {\em J.Phys.} {\bf A42} (2009) 504003,
  [\href{http://xxx.lanl.gov/abs/0906.1663}{{\tt arXiv:0906.1663}}].

\bibitem{Casini:2009sr}
H.~Casini and M.~Huerta, {\it {Entanglement Entropy in Free Quantum Field
  Theory}},  {\em J. Phys.} {\bf A42} (2009) 504007,
  [\href{http://xxx.lanl.gov/abs/0905.2562}{{\tt arXiv:0905.2562}}].

\bibitem{Nishioka:2009un}
T.~Nishioka, S.~Ryu, and T.~Takayanagi, {\it {Holographic Entanglement Entropy:
  an Overview}},  {\em J. Phys.} {\bf A42} (2009) 504008,
  [\href{http://xxx.lanl.gov/abs/0905.0932}{{\tt arXiv:0905.0932}}].

\bibitem{Takayanagi:2012kg}
T.~Takayanagi, {\it {Entanglement Entropy from a Holographic Viewpoint}},  {\em
  Class.Quant.Grav.} {\bf 29} (2012) 153001,
  [\href{http://xxx.lanl.gov/abs/1204.2450}{{\tt arXiv:1204.2450}}].

\bibitem{Herzog:2012bw}
C.~P. Herzog and M.~Spillane, {\it {Tracing Through Scalar Entanglement}},
  \href{http://xxx.lanl.gov/abs/1209.6368}{{\tt arXiv:1209.6368}}.

\bibitem{Casini:2005rm}
H.~Casini, C.~D. Fosco, and M.~Huerta, {\it {Entanglement and Alpha Entropies
  for a Massive Dirac Field in Two Dimensions}},  {\em J. Stat. Mech.} {\bf
  0507} (2005) P07007, [\href{http://xxx.lanl.gov/abs/cond-mat/0505563}{{\tt
  cond-mat/0505563}}].

\bibitem{Calabrese:2004eu}
P.~Calabrese and J.~L. Cardy, {\it {Entanglement Entropy and Quantum Field
  Theory}},  {\em J. Stat. Mech.} {\bf 0406} (2004) P06002,
  [\href{http://xxx.lanl.gov/abs/hep-th/0405152}{{\tt hep-th/0405152}}].

\bibitem{Cardy:2007mb}
J.~Cardy, O.~Castro-Alvaredo, and B.~Doyon, {\it {Form Factors of Branch-Point
  Twist Fields in Quantum Integrable Models and Entanglement Entropy}},  {\em
  J.Statist.Phys.} {\bf 130} (2008) 129--168,
  [\href{http://xxx.lanl.gov/abs/0706.3384}{{\tt arXiv:0706.3384}}].

\bibitem{Doyon:2008vu}
B.~Doyon, {\it {Bi-Partite Entanglement Entropy in Massive Two-Dimensional
  Quantum Field Theory}},  {\em Phys.Rev.Lett.} {\bf 102} (2009) 031602,
  [\href{http://xxx.lanl.gov/abs/0803.1999}{{\tt arXiv:0803.1999}}].

\bibitem{Casini:2009vk}
H.~Casini and M.~Huerta, {\it {Reduced Density Matrix and Internal Dynamics for
  Multicomponent Regions}},  {\em Class.Quant.Grav.} {\bf 26} (2009) 185005,
  [\href{http://xxx.lanl.gov/abs/0903.5284}{{\tt arXiv:0903.5284}}].

\bibitem{Azeyanagi:2007bj}
T.~Azeyanagi, T.~Nishioka, and T.~Takayanagi, {\it {Near Extremal Black Hole
  Entropy as Entanglement Entropy via AdS$_2$/CFT$_1$}},  {\em Phys.Rev.} {\bf
  D77} (2008) 064005, [\href{http://xxx.lanl.gov/abs/0710.2956}{{\tt
  arXiv:0710.2956}}].

\bibitem{Ogawa:2011bz}
N.~Ogawa, T.~Takayanagi, and T.~Ugajin, {\it {Holographic Fermi Surfaces and
  Entanglement Entropy}},  {\em JHEP} {\bf 1201} (2012) 125,
  [\href{http://xxx.lanl.gov/abs/1111.1023}{{\tt arXiv:1111.1023}}].

\bibitem{Coleman:1974bu}
S.~R. Coleman, {\it {The Quantum Sine-Gordon Equation as the Massive Thirring
  Model}},  {\em Phys.Rev.} {\bf D11} (1975) 2088.

\bibitem{Mandelstam:1975hb}
S.~Mandelstam, {\it {Soliton Operators for the Quantized Sine-Gordon
  Equation}},  {\em Phys.Rev.} {\bf D11} (1975) 3026.

\bibitem{DiFrancesco:1997nk}
P.~Di~Francesco, P.~Mathieu, and D.~Senechal, {\it {Conformal Field Theory}}, .
  New York, USA: Springer (1997) 890 p.

\bibitem{Headrick:2010zt}
M.~Headrick, {\it {Entanglement Renyi Entropies in Holographic Theories}},
  {\em Phys.Rev.} {\bf D82} (2010) 126010,
  [\href{http://xxx.lanl.gov/abs/1006.0047}{{\tt arXiv:1006.0047}}].

\bibitem{Fischler:2012uv}
W.~Fischler, A.~Kundu, and S.~Kundu, {\it {Holographic Mutual Information at
  Finite Temperature}},  \href{http://xxx.lanl.gov/abs/1212.4764}{{\tt
  arXiv:1212.4764}}.

\bibitem{Zamolodchikov:1995xk}
A.~B. Zamolodchikov, {\it {Mass Scale in the Sine-Gordon Model and Its
  Reductions}},  {\em Int.J.Mod.Phys.} {\bf A10} (1995) 1125--1150.

\bibitem{Lukyanov:1996jj}
S.~L. Lukyanov and A.~B. Zamolodchikov, {\it {Exact Expectation Values of Local
  Fields in Quantum Sine-Gordon Model}},  {\em Nucl.Phys.} {\bf B493} (1997)
  571--587, [\href{http://xxx.lanl.gov/abs/hep-th/9611238}{{\tt
  hep-th/9611238}}].

\bibitem{Bernard:1994re}
D.~Bernard and A.~LeClair, {\it {Differential Equations for Sine-Gordon
  Correlation Functions at the Free Fermion Point}},  {\em Nucl.Phys.} {\bf
  B426} (1994) 534--558, [\href{http://xxx.lanl.gov/abs/hep-th/9402144}{{\tt
  hep-th/9402144}}].

\bibitem{jin2004quantum}
B.~Jin and V.~Korepin, {\it {Quantum Spin Chain, Toeplitz Determinants and the
  Fisher-Hartwig Conjecture}},  {\em Journal of statistical physics} {\bf 116}
  (2004), no.~1 79--95, [\href{http://xxx.lanl.gov/abs/quant-ph/0304108}{{\tt
  quant-ph/0304108}}].

\bibitem{Its2009fisher}
A.~Its and V.~Korepin, {\it {The Fisher-Hartwig formula and entanglement
  entropy}},  {\em Journal of Statistical Physics} {\bf 137} (2009), no.~5
  1014--1039, [\href{http://xxx.lanl.gov/abs/0906.4511}{{\tt
  arXiv:0906.4511}}].

\bibitem{Witten:1998zw}
E.~Witten, {\it {Anti-de~Sitter Space, Thermal Phase Transition, and
  Confinement in Gauge Theories}},  {\em Adv. Theor. Math. Phys.} {\bf 2}
  (1998) 505--532, [\href{http://xxx.lanl.gov/abs/hep-th/9803131}{{\tt
  hep-th/9803131}}].

\bibitem{Nishioka:2006gr}
T.~Nishioka and T.~Takayanagi, {\it {AdS Bubbles, Entropy and Closed String
  Tachyons}},  {\em JHEP} {\bf 01} (2007) 090,
  [\href{http://xxx.lanl.gov/abs/hep-th/0611035}{{\tt hep-th/0611035}}].

\bibitem{Klebanov:2007ws}
I.~R. Klebanov, D.~Kutasov, and A.~Murugan, {\it {Entanglement as a Probe of
  Confinement}},  {\em Nucl. Phys.} {\bf B796} (2008) 274--293,
  [\href{http://xxx.lanl.gov/abs/0709.2140}{{\tt arXiv:0709.2140}}].

\bibitem{Pakman:2008ui}
A.~Pakman and A.~Parnachev, {\it {Topological Entanglement Entropy and
  Holography}},  {\em JHEP} {\bf 07} (2008) 097,
  [\href{http://xxx.lanl.gov/abs/0805.1891}{{\tt arXiv:0805.1891}}].

\end{thebibliography}\endgroup

\end{document}